\documentclass{elsarticle}

\usepackage{xcolor}
\usepackage{graphicx}
\usepackage{caption}
\usepackage{subcaption}

\newcommand{\R}{I\hspace{-1.2mm}R}

\begin{document}

\begin{frontmatter}
\title{Modified Verhulst-Solow model for long-term population and economic growths}

\author[Gleria]{Iram Gleria\corref{cor1}}
\ead{iram@fis.ufal.br}
\author[Sergio,Sergio1]{Sergio DaSilva}
\author[brenig]{Leon Brenig}
\author[Rocha]{Tarc\'\i sio M.\ Rocha Filho}
\author[Rocha]{Annibal Figueiredo}
\cortext[cor1]{Corresponding author}
\address[Gleria]{Institute of Physics, Federal University of Alagoas, 57072-970 Macei\'o-AL, Brazil.}
\address[Sergio] {Department of Economics, Federal University of Santa Catarina, Florianopolis SC, 88040-900, Brazil.}
\address[Sergio1]{Graduate Program In Economics, Federal University of Espirito Santo, Vitoria ES, 29075-910, Brazil.}
\address[brenig]{Facult\'e des Sciences, Universit\'e Libre de Bruxelles, 1050 Brussels, Belgium.}
\address[Rocha]{Institute of Physics, University of Bras\'\i{}lia, 70919-970 Bras\'\i{}lia-DF, Brazil.}
\begin{abstract}
In this study, we analyze the relationship between human population growth
and economic dynamics. To do so, we present a modied version of the Verhulst
model and the Solow model, which together simulate population dynamics and
the role of economic variables in capital accumulation. The model incorporates
support and foraging functions, which participate in the dynamic relationship
between population growth and the creation and destruction of carrying capacity. The validity of the model is demonstrated using empirical data.
\end{abstract}
\end{frontmatter}


\section{Introduction}

The main objective of this work is to develop and test a model to explain the ``quasi-permanent'' growth of the human population in the last $12$ thousand years. For ``quasi-permanent'' growth
we understand the fact that throughout this period the growth rates of the human population were mostly positive, as we can see in figure \ref{fig8} below.

The carefull analysis of figure 1 shows that the world’s population has been growing continuously from 10000BC to 0, with exponential growth rates that vary little but always seem to increase until the year 0, with the exception of the interval 2000BC-1000BC, which appears to have had slightly lower growth than the previous millennium and a strong recovery in the following millennium.
When we compare the period 10000BC-0 to the subsequent period, we can clearly see that the world experienced a significant decrease in population growth rate between the years 0 and 1000, recovered later, and experienced an unprecedented growth inflection from 1700 onwards. 

Also, figure \ref{fig8} depicts the appearance of negative growth rates, for short time periods, between the years 0 and 1500. It is impossible to say categorically that the appearance of negative rates is unique to this period and did not occur in the previous period of 10000BC-0, because population measurements in the latter are made at 1000-year intervals. In any case, the millennium following the year 0 was clearly a period of lower growth of the world population when compared to the period 10000BC-0.
Another important point to note in Figure \ref{fig8} is that fluctuations in population growth rates after the year 0  occur significantly in periods of 50 to 100 years.

\begin{figure}
[!htb]
\begin{center}
\hspace*{-4mm}\includegraphics[width= 6.1cm]{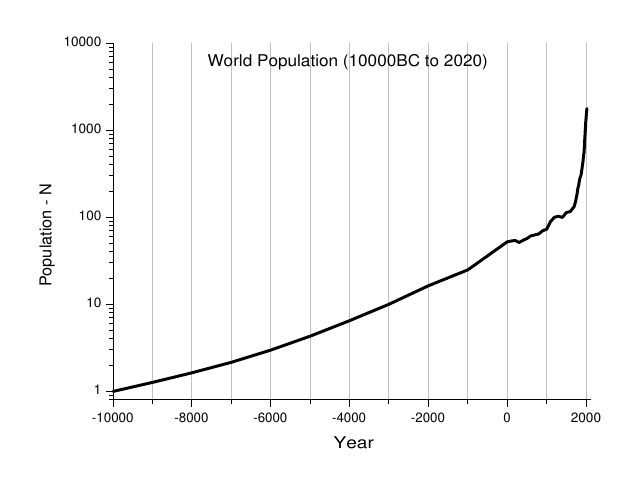}
\hspace*{-6mm}\includegraphics[width= 6.9cm]{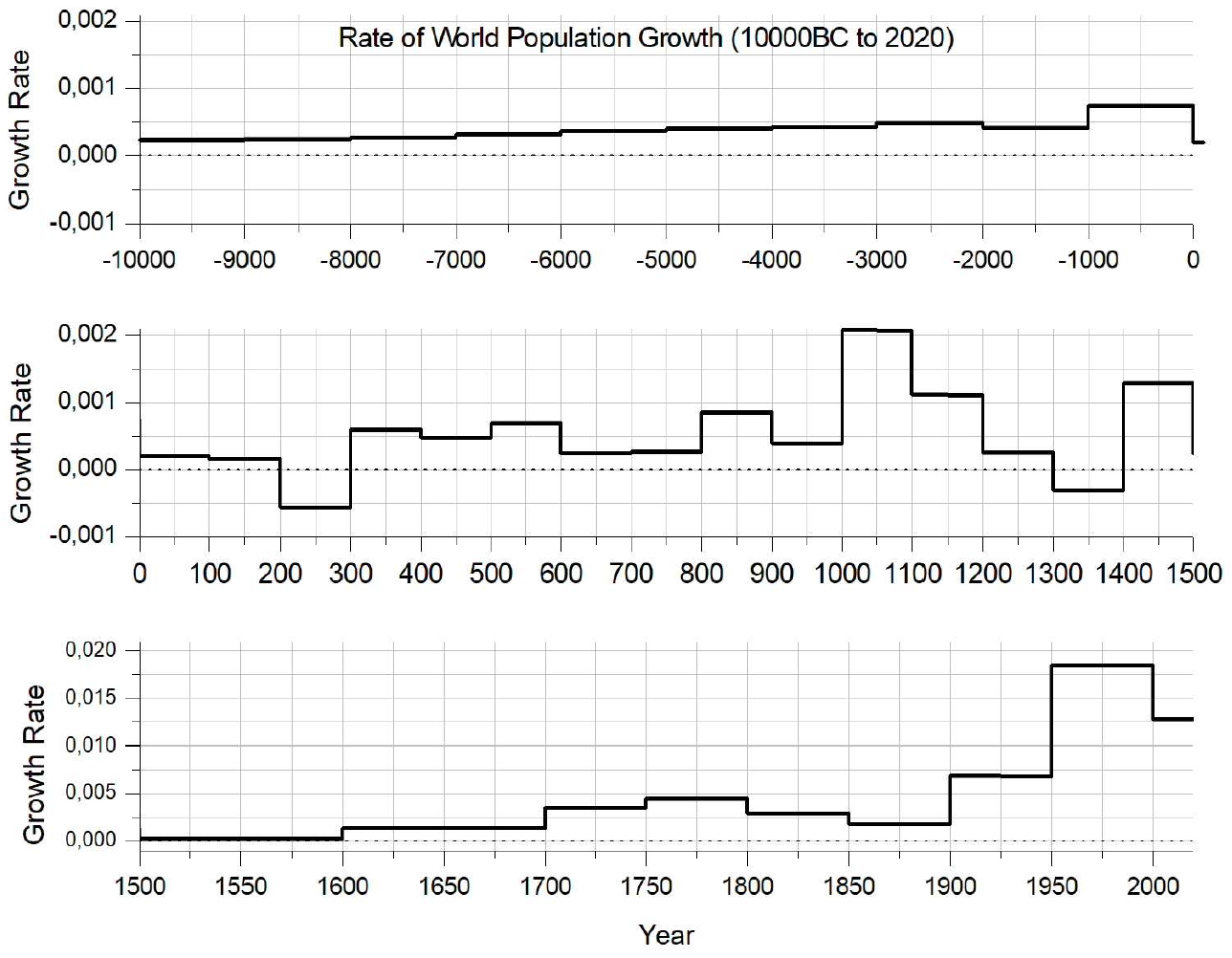}
\end{center}
\vspace*{-4mm}
\caption{The data for population can be found in Ref. \cite{population}. The variables are given as a proportion of their respective values in the initial reference year 10000BC (denoted as -10000 in the figure). The vertical axis is on a monolog scale.
The growth rate is calculated as $(t_{n+1}-t_n)^{-1}\ln[N(t_{n+1})/N(t_n)]$, where $t_n$ and $t_{n+1}$ are consecutive time periods during which the population $N$ is measured.}
\label{fig8}
\end{figure}

To explain this pattern of ``quasi-permanent'' growth of the human population and, correspondingly, its economic growth, we developed a model based on the coupling of two equations: the Velhurst equation to describe population growth of an animal species and the Solow equation for capital growth. Moreover, it is added to these two equations an equation for the creation and loss of carrying capacity.
The main characteristic of our model is to take into account that the process of competition for resources (carrying capacity), as modeled in the Velhurst equation and which make possible to maintain a certain level of population, is intermediated by a process of production of wealth that will be determined by a production function dependent on two productive factors: capital and labor.

The Solow model, through a production function, predicts aggregate output growth based on changes in the population growth rate, the savings rate, and the rate of technological progress~\cite{solow,swan}. Economic growth is exogenous because the population grows at a constant rate, consumers save a constant proportion of their incomes, and firms produce output using the same production technology that takes in capital and labor as inputs. Because the production function has constant returns-to-scale, the Solow growth model focuses on output and capital per worker rather than aggregate output and capital stock. A capital accumulation equation expresses the relationships between present and future capital stock, capital depreciation rate, and capital investment level. As a result, there is no long-run growth. Furthermore, capital accumulation implies increased output per worker only if it is accompanied by improved technology. As a result, sustained growth requires technological progress. An increase in the population growth rate, in particular, raises the growth rate of aggregate output but not the growth rate of output per worker. (See Ref. ~\cite{nova1} for a primer on the Solow model and its main extensions.)

The most well-known Solow model extension incorporates human capital to explain the lack of observed convergence ~\cite{nova2}. Countries are wealthy due to high rates of investment, not only in physical capital but also in human capital, low population growth rates, and high levels of technology. In the steady state, however, output per worker grows at the same rate as technological progress, as in the original Solow model. This means that the dynamics are essentially the same as the basic model. Other generalizations of the Solow model include a non-constant labor growth rate, in which countries with equal capital per capita but different labor growth rates end up with the same capital per worker for similar limits of labor growth rates \cite{guerrini}; population discrimination between employed, unemployed, and economically inactive with time-delay inducing oscillations in the active population and capital stock \cite{kaddar}; a single country’s specific demographic growth pattern \cite{wang}; and a labor force that follows Malthusian law and has a constant migration rate \cite{neto}. 

We are not directly interested in most aspects of the Solow model discussed in the two previous paragraphs, since all of them are related to the imposition of a production function with constant returns-to-scale, implying population dynamics and economic growth exogenous to the model.  
Indeed, in our model, Capital and Population are variables of a coupled system of ODE's and their respective dynamics are endogenous and directly related to the process of creation and loss of carrying capacity.

An important modification, with respect to the Solow model, is not to assume that the Cobb-Douglas production function has constant returns-to-scale. This fact will allow us to show that the enrichment (or impoverishment) of population will be directly related to the returns-to-scale of the production function. Therefore, population growth, accompanied by enrichment or impoverishment, will be strictly related to greater production of wealth using the same amount of carrying capacity.

Applying the model developed to explain the real evolution of the human population, we consider that the wealth produced is measured by standardized measures of Gross National Product (GNP) and the carrying capacity will be measured as the total amount of energy consumed by the population in a certain period. In fact, the consumed energy can be seen as an approximate and standardized measure of the amount of resources used in the process of wealth production. The results presented in this work suggest that the choice of this measure for the carrying capacity seems adjusted and adequate to explain the evolution of the human population in the last $200$ years.  

The structure of this paper is as follows. Section \ref{putz} presents our model. Section \ref{finite} takes into account the finite carrying capacity, which limits population growth. Section \ref{infinite} considers the possibility of infinite carriyng capacity limit. Section \ref{figuras} illustrates it with historical real-world data, Section \ref{exponential} considers the exponential growth of carrying capacity limit, and Section \ref{fim} concludes.  In the appendices we presents some important analytical findings.

\section{The model}
\label{putz}
Let us consider a capital stock $K$ with production function $Y(K,L)$, with $L$ the labor input~\cite{solow,swan}. Then, the equation for capital growth is given by
\begin{equation}
\label{int1}
\frac{dK}{dt}=s_0Y(K,L)-dK,\; s_0>0,\; d>0,
\end{equation}
where $s_0,d>0$ are constant parameters, with $d$ the capital depreciation (rusty machines, broken tools, shattered roads),
and $s_0$ the capital accumulation rate relative to the production function $Y(K,L)$.  We assume  that the labor $L$ is a linear function of the entire population $N$, that is, $L=z_0N$ with $z_0>0$.

In our generalized model we make a modification of the Verhulst model~\cite{logis} for the population $N$. We consider:
\begin{equation}
\label{int2}
\frac{dN}{dt}=N\left[b-r_0\frac{S(K,L)}{R}\right],\;b>0,\; r_0>0,
\end{equation}
where $b>0$ is the population growth rate, and $r_0$ represents the competition parameter for the carrying capacity $R$.
Equation~(\ref{int2}) differs from the usual Verhulst model as the competitive term
is not a function the population size $N$ but a function of the capital $K$ and workforce $L$ through a given
support function $S(K,L)$, which is supposed to be a homogeneous function of first degree. The rate of change for carrying capacity $R$ is considered to depend on labor and capital through the differential equation:
\begin{equation}
\label{int3}
\frac{dR}{dt}=-e_0S(K,L)+g_0S(K,L)\left(1-\frac{R}{R_T}\right),\;
e_0>0,\;g_0> 0,\;R_T>0,
\end{equation}
where $R_T$ denotes the carrying capacity limit, $-e_0S(K,L)$ represents the inevitable loss of carrying capacity,
with $e_0$ the loss rate per support unit. The second term $g_0S(K,L)(1-{R}/{R_T})$ denotes an increase in the carrying capacity, coming for
instance from a more efficient use of commodities, and the support function $S(K,L)$ plays the role of a foraging function.
The complete model is then expressed by the following set of coupled ordinary differential equations:
\begin{eqnarray}
\label{int5}
\frac{dK}{dt}&=&-dK+sY(K,z_0N), \nonumber\\
\frac{dN}{dt}&=&N\left[b-r_0\frac{S(K,z_0N)}{R}\right],\\ 
\frac{dR}{dt}&=&-e_0S(K,z_0N)+g_0S(K,z_0N)\left(1-\frac{R}{R_T}\right),\nonumber
\end{eqnarray}
where we have taken into account the relation $L=z_0N$ in equations (\ref{int1}), (\ref{int2}) and (\ref{int3}).

In order to proceed with our approach, we assume that the functions $Y$ and $S$ are of the form
\begin{equation}
\label{int6}
	Y=K^{\alpha_1}L^{\alpha_2}=z_0^{\alpha_2}K^{\alpha_1}N^{\alpha_2};\hspace{3mm}S=Y^{\beta}=\left(K^{\alpha_1}L^{\alpha_2}\right)^{\beta}=z_0^{\alpha_2\beta}K^{\alpha_1\beta}N^{\alpha_2\beta},
\end{equation}
where 
\begin{equation}
\label{cond_expoentes}
0<\alpha_1<1,\;\;\alpha_2>0,\;\;\beta> 0,\;\; (\alpha_1+\alpha_2)\beta=1.
\end{equation}

Therefore, Eq.~(\ref{int5}) becomes a Quasi-Polynomial (QP) system~\cite{8} and may be written as
\begin{equation}
\label{qp1}
\stackrel{\cdot }{x}_i=c_i x_i+x_i\sum_{j=1}^m A_{ij}\prod_{k=1}^n
x_k^{B_{jk}};\;\; i=1,\ldots,n,\;x_i\in\R.
\end{equation}
From (\ref{qp1}) we define a $n\times m$ matrix ${\bf A}$ with elements $A_{ij}$ and also the $m\times n$ matrix ${\bf B}$ with 
elements $B_{jk}$, where $m$ is the number of different quasi-monomials and $n$ the number of variables.
The term {\it quasi-monomials} comes from the possibility on non integer values for the entries of matrix $B$.
Defining the embedding defined by
\begin{equation}
	U_i=\prod_{k=1}^n x_k^{B_{ik}};\hspace{3mm} i=1,\ldots,m,
	\label{defx}
\end{equation}
the QP system in Eq.~(\ref{qp1}) is recast in a $m$-dimensional quadratic Lotka-Volterra system of type:
\begin{equation}
\label{LVdef}
\dot{U}_i=U_i\left(l_i+\sum_{i=1}^m M_{ij}U_j\right),
\end{equation}
where $l_i=\sum_j B_{ij}c_j$ and $M_{ij}$ are the components of the $m\times m$ matrix ${\bf M=BA}$.
A number of analytical properties ensues from the LV-format and will be used in what follows~\cite{jmph}-\cite{nos6}.

We now turn to analyze the Lotka-Volterra format in two specific situations: finite and infinite limit
for carrying capacity.

\section{Finite carrying capacity limit}
\label{finite}

For a finite carrying capacity limit $R_T$, it is helpful to divide both sides of (\ref{int5}) by the depreciation rate $d$,
and perform the time reparametrization $t\rightarrow t'/d$ to obtain:
\begin{eqnarray}
\label{A4}
\frac{dK}{dt}&=&K\left(-1+\frac{s}{d}K^{\alpha_1-1}N^{\alpha_2}\right),\nonumber\\
\frac{dN}{dt}&=&N\left(\displaystyle{\frac{b}{d}}-\frac{r}{d}\frac{K^{\beta\alpha_1}N^{\beta\alpha_2}}{R}\right),\\
\frac{dR}{dt}&=&R\left(\frac{g-e}{d}\frac{K^{\beta\alpha_1}N^{\beta\alpha_2}}{R}-\frac{g}{dR_T}K^{\beta\alpha_1}N^{\beta\alpha_2}\right),\nonumber
\end{eqnarray}
where at the end we changed back to the original variable, $t'\Leftrightarrow t$. 
The parameters appearing in (\ref{A4}) are given by:
\begin{equation}
\label{novos_parametros}
s=s_0z_0^{\alpha_2},\;\;r=r_0z^{\beta\alpha_2},\;\;e=e_0z^{\beta\alpha_2},\;\;g=g_0z^{\beta\alpha_2}.
\end{equation}

Equation~(\ref{A4}) is a well defined QP system, and their corresponding quasi-monomials are given by
\begin{equation}
\label{monomiosU}
U_1=K^{\alpha_1-1}N^{\alpha_2},\; U_2=K^{\beta\alpha_1}N^{\beta\alpha_2}R^{-1},\; U_3=K^{\beta\alpha_1}N^{\beta\alpha_2}.
\end{equation}
Let us now define the rescaled monomial variables
\begin{equation}
\label{monomiosX}
X_1=\frac{s}{d}U_1,\;\; X_2=\frac{r}{d}U_2,\;\; X_3=\frac{g}{R_Td}U_3,
\end{equation} 
we obtain from the Lotka-Volterra system (\ref{LVdef}) a simpler system written as
\begin{equation}
\label{A18}
\frac{dX_i}{dt}=X_i\left({\bar l}_i+\sum_{i=1}^3{\bar M}_{ij}X_j\right),
\end{equation}
where ${\bf \bar l}$ and ${\bf\bar M}$ are
\begin{equation}
\label{A19}
	{\bf\bar l}=(\bar l_i)=\left(\begin{array}{c}
\displaystyle 1-\alpha_1+\displaystyle{\bar{b}}\alpha_2 \vspace*{2mm}\\
\displaystyle \beta\left(-\alpha_1+\displaystyle{\bar{b}}\alpha_2\right)\vspace*{2mm}\\
\displaystyle \beta\left(-\alpha_1+\displaystyle{\bar{b}}\alpha_2\right)\vspace*{2mm}
\end{array}\right),\;
	{\bf \bar M}=(\bar M_{ij})=\left(\begin{array}{ccc}
\alpha_1-1 & -\alpha_2 & 0 \vspace*{2mm}\\
\beta\alpha_1 & -\bar{k} -\beta\alpha_2 & 1 
\vspace*{2mm}\\ 
\displaystyle \beta\alpha_1 & 
\displaystyle -\beta\alpha_2 & 0
\end{array}\right).  
\end{equation}
where
\begin{equation}
\label{parametros_bk}
\bar{b}\equiv\frac{b}{d},\;\;\bar{k}\equiv \frac{g-e}{r}=\frac{g_0-e_0}{r_0}.
\end{equation}

The explicit system reads:
\begin{eqnarray}
\label{eq2}
	\frac{dX_1}{dt} & = & X_1\left(1-\alpha_1+\displaystyle{\bar{b}}\alpha_2-(1-\alpha_1)X_1-\alpha_2X_2\right),
	\nonumber\\
	\frac{dX_2}{dt} & = &  X_2\left(-\beta\alpha_1+\displaystyle{\bar{b}}\beta\alpha_2+\beta\alpha_1X_1-(\bar{k}+\beta\alpha_2)X_2+X_3\right),
	\\
	\frac{dX_3}{dt} & = &  X_3\left(-\beta\alpha_1+\displaystyle{\bar{b}}\beta\alpha_2+\beta\alpha_1X_1-\beta\alpha_2X_2\right).\nonumber
\end{eqnarray}
The growth rates of $K$, $N$ and $R$ are the variables of interest, and are determined from equation (\ref{A4}) using the definitions of variables  $U_i$ in Eq.~(\ref{monomiosU}) and $X_i$ in Eq.~(\ref{monomiosX}):
\begin{eqnarray}
\label{new0}
	\lambda_K & \equiv & \frac{1}{K}\frac{dK}{dt}=-1+X_1,\nonumber\\
	\lambda_N & \equiv & \frac{1}{N}\frac{dN}{dt}=\bar{b}-X_2, \\
	\lambda_R & \equiv & \frac{1}{R}\frac{dR}{dt}=\bar{k} X_2-X_3.\nonumber
\end{eqnarray}

Since  ${\bar k}>0$, there is a stationary point of Eq. (\ref{eq2}) that belongs to the interior of the positive orthant ${\R}^3_+$, and it is given by 
\begin{equation}
\label{A23}
X_1^*=1,\;\;X_2^*={\bar b},\;\;X_3^*={\bar k}\bar{b}.
\end{equation}
Let us remember that the positive orthant ${\R}^3_+$ is defined as the set of vectors belonging to $\R^3$ such that theirs components are positive. The eigenvalues of the Jacobian matrix of this fixed point~(\ref{A23}) is given by
\begin{equation}
\label{A25}
\begin{array}{l}
\lambda_1^*=\displaystyle -\frac{\Delta_1}{2}-\frac{1}{2}\sqrt{\Delta_1^2-4{\bar b}\beta\alpha_2},\\ \\
\lambda_2^*=\displaystyle -\frac{\Delta_1}{2}+\frac{1}{2}\sqrt{\Delta_1^2-4{\bar  b}\beta\alpha_2},\\
\end{array}\;\;\;
\lambda_3^*= -{\bar b}{\bar k},
\end{equation}
with 
\begin{equation}
\label{delta}
\Delta_1=(1-\alpha_1)+{\bar b}\beta_2,\;\; \beta\alpha_2=-\det({\bf B})=-\det({\bf \bar M})=\beta\alpha_2.
\end{equation}
As ${\bar k}>0$ and then $\lambda^*_3<0$, implying a stable manifold associated with this eigenvalue. Therefore, the stability of the fixed point (\ref{A23}) will depend only on the analysis of the eigenvalues $\lambda^*_1$ and $\lambda^*_2$. 
As $\beta\alpha_2>0$ and $\Delta_1>0$ then $\lambda^*_1$ and $\lambda^*_2$ have a negative real part. Therefore, the fixed point (\ref{A23}) is locally stable.
Furthermore, we can show (see Appendix A) that this fixed point is globally stable, implying that any solution of the system (\ref{eq2}), with initial condition in the positive orthant, will converge asymptotically ($t\rightarrow \infty$) to this fixed point.

Using the Quasi-Monomial transformations (\ref{monomiosU}) and their respective linear rescaling (\ref{monomiosX}), the global stability property of the fixed point (\ref{A23}) can be expressed in terms of the variables $K$ , $N$ and $R$, that is, for ${\bar k}>0$ any solution  with initial condition such that $K(0)>0$, $L(0)>0$ and $R(0)>0$ will converge asymptotically to a single well-defined fixed point of the QP system in (\ref{A4}) :
\begin{eqnarray}
\label{pfklr}
 \lim_{t\rightarrow\infty}K&=&K^*=\left(\frac{b}{r}R^*\right)^{\beta^{-1}}\left(\frac{s}{d}\right),\nonumber \\
 \lim_{t\rightarrow\infty}N&=&N^*=\left(\frac{b}{r}R^*\right)^{\frac{1-\alpha_1}{\beta\alpha_2}}\left(\frac{d}{s}\right)^{\frac{\alpha_1}{\alpha_2}}, \\
\lim_{t\rightarrow\infty}R&=&R^*=\frac{g-e}{g}R_T=\frac{g_0-e_0}{g_0}R_T,\nonumber
\end{eqnarray}

It is worth noting that $\beta\alpha_2=-\det({\bf B})>0$ leads to the existence of an inverse transformation of the Quasi-Monomial transformation in (\ref{monomiosU}). This inverse transformation will be a Quasi-Monomial transformation defined by the inverse of the matrix ${\bf B}$.

The main characteristic of the differential model in (\ref{A4}) is that for any initial condition the variables $K$, $N$ and $R$ will saturate in a finite value, that is, asymptotically all solutions of the model will converge to a stagnant solution:
\begin{equation}
\label{estagnante}
\lim_{t\rightarrow\infty}\lambda_K= \lim_{t\rightarrow\infty}\lambda_N=\lim_{t\rightarrow\infty}\lambda_R=0.
\end{equation}

This property is a clear consequence of having included a Velhurst equation that endogenizes the variable associated with the population (or the productive factor associated with work). Indeed, this equation implies competition for carrying capacity between the capital and population. On the other hand, as the resource growth capacity is limited  ($R_T<\infty$), it becomes impossible to have a permanent growth of these factors.

We can clearly see in Eq. (\ref{pfklr}) that the variables $K$ and $N$ saturate in a value that is directly proportional to some positive power of the variable $R^*$. On the other hand, the saturation value $R^*$ of the variable $R$ is directly and linearly proportional to the parameter $R_T$. This fact perfectly justifies the interpretation of the parameter $R_T$ as a measure of a limit for the carrying capacity $R$.

Two per capita variables are very important: the capital per population and  the output per population. These two relative variables calculated for the stationary solution in (\ref{pfklr}) are given by
\begin{eqnarray}
\label{razaokl}
\frac{K^*}{N^*}&=&\left(\frac{b}{r}R^*\right)^{\frac{\alpha_1+\alpha_2-1}{\beta\alpha_2}}\left(\frac{s}{d}\right)^{\frac{\alpha_1+\alpha_2}{\alpha_2}},\nonumber\\
\frac{Y^*}{N^*}&=&\left(\frac{b}{r}R^*\right)^{\frac{\alpha_1+\alpha_2-1}{\beta\alpha_2}}\left(\frac{s}{d}\right)^{\frac{\alpha_1}{\alpha_2}},
\end{eqnarray}
where $Y^*={K^*}^{\alpha_1}{N^*}^{\alpha_2}$.

Incidentally, for $\alpha_1+\alpha_2=1$, the capital per population at the fixed point is given by
\begin{equation}
\label{pfcobb}
\frac{K^*}{N^*}=\left(\frac{s}{d}\right)^{\frac{1}{1-\alpha_1}},\;\;\frac{Y^*}{N^*}=\left(\frac{s}{d}\right)^{\frac{\alpha_1}{(1-\alpha_1)}}.
\end{equation}
In this case, we see that these per capita relationships do not depend on the level of carrying capacity $R^*$,
so the increase in population due to an increase in carrying capacity does not imply either enrichment or impoverishment per capita of the population, which is typical of the Solow model, with the significant difference that there is no economic growth here, as the asymptotic solutions are stagnant at fixed values of capital and population.

\section{Infinite carrying capacity limit}
\label{infinite}

Let us now analyze the case where the limit for carrying capacity is  infinite, {\it i.e.} $R_T=\infty$. In this case we have to consider $X_3=0$ in the system (\ref{A18}), which leads us to the following two-dimensional Lotka-Volterra system:
\begin{eqnarray}
\label{new2}
\frac{dX_1}{dt}&=&X_1\left(1-\alpha_1+{\bar b}\alpha_2-(1-\alpha_1)X_1-\alpha_2X_2\right),\nonumber\\
\frac{dX_2}{dt}&=&X_2\left(-\beta\alpha_1+{\bar b}\beta\alpha_2+\beta\alpha_1X_1-({\bar k}+\beta\alpha_2)X_2\right),
\end{eqnarray}
In this case, (\ref{new0}) assumes the form
\begin{eqnarray}
\label{new9}
\lambda_K=\frac{1}{K}\frac{dK}{dt}=-1+X_1,\;\lambda_N=\frac{1}{N}\frac{dN}{dt}={\bar b}-X_2,\;\lambda_R=\frac{1}{R}\frac{dR}{dt}={\bar k}X_2.
\end{eqnarray}
The system in (\ref{new2}) has the following fixed point:
\begin{equation}
\label{new3}
X_1^{**}=\frac{\beta\alpha_2+{\bar k}\left(1-\alpha_1+{\bar b}\alpha_2\right)}{\beta\alpha_2+{\bar k}(1-\alpha_1)},\;\;X_2^{**}=
\frac{{\bar b}\beta\alpha_2}{\beta\alpha_2+{\bar k}(1-\alpha_1)},
\end{equation}
This fixed point represents a stationary solution of system (\ref{new2}) and  
$$ {\bar k}>-\frac{\beta\alpha_2}{1-\alpha_1+{\bar b}\alpha_2}\;  \Rightarrow \; X_1^{**}>0,\; X_2 ^{**}>0.$$
Moreover, in appendix A is shown the existence and global stability of the fixed point (\ref{new3}) for any value of ${\bar k}$ such that 
\begin{equation}
\label{condicaoee}
{\bar k}>\max\left\{-\beta_1,-\frac{\beta\alpha_2}{1-\alpha_1+{\bar b}\alpha_2}\right\}
\end{equation}
Therefore, the fixed point in Eq.~(\ref{new3}) becomes a global attractor for all solutions with
strictly positive initial conditions $K(0)>0$, $N(0)>0$ and $R(0)>0$. 

Substituting the expressions of this fixed point in Eq. (\ref{new9}) and considering the original time variable of the model (\ref{A4}) through the inverse reparametrization $t \rightarrow d\,t'$ 
we obtain
\begin{equation}
\label{new8a}
	\lambda_K=\frac{\alpha_2{b}{\bar k}}{(1-\alpha_1){\bar k}+\beta\alpha_2},
\;\lambda_N=\frac{(1-\alpha_1){b}{\bar k}}{(1-\alpha_1){\bar k}+\beta\alpha_2},
	\;\lambda_R=\frac{{b}{\bar k}\beta\alpha_2}{(1-\alpha_1){\bar k}+\beta\alpha_2},
\end{equation}
which shows that the stationary solution (\ref{new3}) of Eq. (\ref{new2}) corresponds to trajectories with exponential growth/degrowth for $K$, $N$, and $R$, whenever ${\bar k}$ satisfies Eq. (\ref{condicaoee}) and 
$\alpha_1<1$, because in this case we certainly have $\lambda_K>0$, $\lambda_N>0$ and $\lambda_R>0$.
This result is consistent with the fact that we expect no boundaries for growth rates in an economy with unlimited carrying capacity
and generalizes the classical Solow model.

Using the relations (\ref{defx}) for $U_1$ and $U_2$ and their respective linear scaling in (\ref{monomiosX}) we have the following asymptotic relations:
\begin{eqnarray}
\label{rlc_solow}
&&\hspace*{-10mm}\lim_{t\rightarrow\infty}K=\left(\frac{b}{r}\frac{\beta\alpha_2}{\beta\alpha_2+{\bar k}(1-\alpha_1)}R\right)^{\beta^{-1}}
\left(\frac{s}{d}\frac{\beta\alpha_2+{\bar k}(1-\alpha_1)}{\beta\alpha_2+{\bar k}\left(1-\alpha_1+{\bar b}\alpha_2\right)}\right)\nonumber\\
&&\hspace*{-10mm}\lim_{t\rightarrow\infty}N=\left(\frac{b}{r}\frac{\beta\alpha_2}{\beta\alpha_2+{\bar k}(1-\alpha_1)}R\right)^{\frac{1-\alpha_1}{\beta\alpha_2}}
\left(\frac{d}{s}\frac{\beta\alpha_2+{\bar k}\left(1-\alpha_1+{\bar b}\alpha_2\right) }{\beta\alpha_2+{\bar k}(1-\alpha_1)}\right)^{\frac{\alpha_1}{\alpha_2}}\\
&&\hspace*{-10mm}\lim_{t\rightarrow\infty}R=R_0\exp\left(\lim_{t\rightarrow\infty}\int_0^t{\bar k}X_2(t')dt'\right)=R_0\exp\left(\frac{{\bar k}{b}\beta\alpha_2}{\beta\alpha_2+{\bar k}(1-\alpha_1)}t \right)\nonumber
\end{eqnarray}
The asymptotic expressions above show that capital and population factors grow as a non-linear scaling proportion of the resource factor, which in turn grows exponentially. We can interpret this asymptotic solution as representing a production process whose growth is dependent on a resource factor that must grow geometrically, thus implying that the capital and population factors must also grow geometrically.
However, we must be careful here to emphasize that we are not saying that the growth of carrying capacity causes the growth of capital and population, because in fact, according to the non-linear model in (\ref{int5}) and $R_T=\infty$, we see that the equation for the variable $R$  depends on the variables $K$ and $N$ through the balancing of the resource loss and  gain processes, which in turn is determined by the support function.

Now, using the expressions in (\ref{rlc_solow}), we can obtain respectively the per capita relations of the capital and output factors with respect to the labor factor:
\begin{eqnarray}
\label{rlc_percapita}
&&\hspace*{-8mm}\displaystyle \lim_{t\rightarrow\infty}\frac{K}{N}=\left(\frac{b}{r}\frac{\beta\alpha_2}{\beta\alpha_2+{\bar k}(1-\alpha_1)}R\right)^{\frac{\alpha_1+\alpha_2-1}{\beta\alpha_2}}
\left(\frac{s}{d}\frac{\beta\alpha_2+{\bar k}(1-\alpha_1)}{\beta\alpha_2+{\bar k}\left(1-\alpha_1+{\bar b}\alpha_2\right)}\right)^{\frac{\alpha_1+\alpha_2}{\alpha_2}} \nonumber\\
&&\hspace*{-8mm}\displaystyle \lim_{t\rightarrow\infty}\frac{Y}{N}=\left(\frac{b}{r}\frac{\beta\alpha_2}{\beta\alpha_2+{\bar k}(1-\alpha_1)}R\right)^{\frac{\alpha_1+\alpha_2-1}{\beta\alpha_2}} 
\left(\frac{s}{d}\frac{\beta\alpha_2+{\bar k}(1-\alpha_1)}{\beta\alpha_2+{\bar k}\left(1-\alpha_1+{\bar b}\alpha_2\right)}\right)^{\frac{\alpha_1}{\alpha_2}}
\end{eqnarray}
It is also interesting to calculate the rate between the growth rates:
\begin{equation}
\label{A33}
\frac{\lambda_K}{\lambda_N}=\frac{\lambda_Y}{\lambda_L}=\frac{\alpha_2}{1-\alpha_1},
\end{equation}
where the output growth rate is given by
\begin{equation}
\label{A34}
\lambda_Y=\frac{1}{Y}\frac{dY}{dt}=\alpha_1\lambda_K+\alpha_2\lambda_N=\lambda_K.
\end{equation}
Finally, from the relation in Eq. (\ref{A33}) and Eq. (\ref{cond_expoentes}) it is straightforward to show that
\begin{eqnarray}
\label{A34a}
\beta^{-1}=\alpha_1+\alpha_2<1 &\Rightarrow& \lambda_K<\lambda_N\;(\lambda_Y<\lambda_N),\nonumber\\
\beta^{-1}=\alpha_1+\alpha_2=1 &\Rightarrow& \lambda_K=\lambda_N\;(\lambda_Y=\lambda_N),\\
\beta^{-1}=\alpha_1+\alpha_2>1 &\Rightarrow& \lambda_K>\lambda_N\;(\lambda_Y>\lambda_N).\nonumber
\end{eqnarray}
The relationships in (\ref{rlc_percapita}) and (\ref{A34a}) imply a very important interpretation for the global scale factor 
$\beta^{-1}=\alpha_1+\alpha_2$ of the production function. 

When $\alpha_1+\alpha_2<1$ (decreasing returns of the production function) we have that the capital and product-to-population ratios depend inversely on the carrying capacity $R$, that is, the higher the level of carrying capacity used in the production process, the smaller the amount of capital and product per population. This type of feature can be interpreted as typical of a Malthusian model, where the increase in population due to an increase in carrying capacity $R$ implies the impoverishment per capita of the population. 

When $\alpha_1+\alpha_2>1$ (which implies increasing returns of the production function) we have that the ratios of capital and product per population depend directly on the carrying capacity $R$, that is, the higher the level of carrying capacity used in the production process, the greater the amount of capital and product per population. This type of feature can be interpreted as typical of a non-Malthusian model, where the increase in  population due to an increase in the carrying capacity $R$ implies the per capita enrichment of the population.

For the case $\alpha_1+\alpha_2=1$ (as in the original Solow Model) we have $\lambda_N=\lambda_K=\lambda_Y$ and
the capital and output  per capita in Eq.~(\ref{rlc_percapita}) becomes:  

\begin{eqnarray}
\label{solow_percapita}
\lim_{t\rightarrow\infty}\frac{K}{N}&=&\left(\frac{s}{d}\frac{\beta\alpha_2+{\bar k}(1-\alpha_1)}{\beta\alpha_2+{\bar k}\left(1-\alpha_1+{\bar b}\alpha_2\right)}\right)^{\frac{1}{1-\alpha_1}} \nonumber\\
\lim_{t\rightarrow\infty}\frac{Y}{N}&=&\left(\frac{s}{d}\frac{\beta\alpha_2+{\bar k}(1-\alpha_1)}{\beta\alpha_2+{\bar k}\left(1-\alpha_1+{\bar b}\alpha_2\right)}\right)^{\frac{\alpha_1}{1-\alpha_1}}.
\end{eqnarray}
Thus, we can clearly see that the capital and and output per capita stay constant while the population and capital grow
exponentially, however, unlike in the classical Solow model, the capital and output  per capita does not depend
uniquely on the ratio $s_0/d$.

It is worth noting that the name Malthusian model is inspired by the original Malthus idea: wealth and population both increase over time, but their per capita ratio decreases. However, here, unlike Malthus's original model, both population and wealth grow geometrically but at different rates.

\section{Applications}
\label{figuras}


The goal of this section is to determine whether the stationary solutions given by equation (\ref{pfklr}) of the finite carrying capacity limit model (\ref{A4}) or, alternatively, the stationary solutions
(\ref{rlc_solow}) of the model with infinite support capacity (\ref{new2}), explain the empirically observed evolution of the variables $N$, $K$, and $R$. To carry out this comparison, we select the empirical variables that correspond to the variables $K$ and $R$ in our models:  the capital variable $K$ is not considered because it is difficult to measure directly, instead, we employ the output variable $Y$ measured from Gross National Product (GNP); we choose the energy consumption as a measure to represent the variable $R$ (carrying capacity), namely the equivalent energy measure corresponding to the quantities of energy sources consumed. It is worth noting that the exponential growth rates of capital $K$ and product $Y$ are equal for the solutions corresponding to the system’s fixed point, according to equations (\ref{A33}) and (\ref{new2}).

We have a problem with the historical periods to be analyzed because the existing estimates for the three empirical variables chosen do not cover the same time intervals. Estimates for the variable $N$ range from 10000BC to 2020AD; estimates for the variable $Y$ range from 0 to 2020AD; and estimates for the variable $R$ range only from 1820 to 2020. We examine this later period in greater depth because it is the only period to have empirical estimates for $N$, $Y$, and $R$. Following that, and in light of previous analysis, we consider the period 0-1820.

\subsection{Period of 1820-2020}

\begin{figure}
[!htb]
\begin{center}
\hspace*{-10mm}\includegraphics[width= 13.7cm]{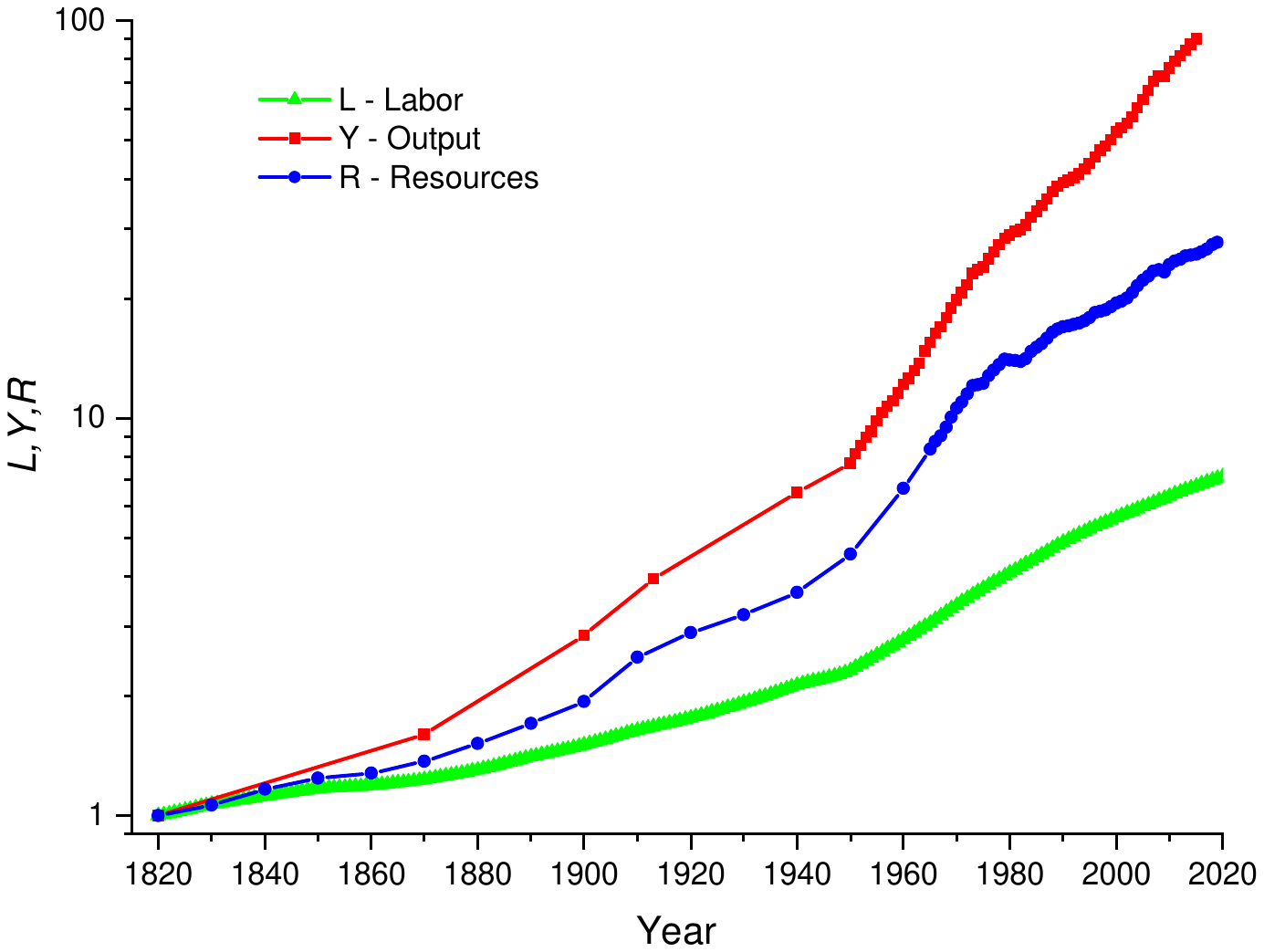}
\end{center}
\vspace*{-8mm}
\caption{Ref. \cite{population} is the source of population data, Ref. \cite{gdp} is the source of output data, and Ref.
\cite{energy} is the source of energy consumption data. All variables are expressed as a percentage of their initial reference year 1820 values. A monolog scale is used for the vertical axis.}
\label{fig1}
\end{figure}

Figure \ref{fig1} depicts the temporal evolution of the variables $N$, $Y$, and $R$ in a monolog scale from 1820 to 2020. We see that the three variables present a permanent growth regime, albeit one that fluctuates over the time period under consideration.

\begin{figure}
[!htb]
\begin{center}
\hspace*{-10mm}\includegraphics[width= 13.7cm]{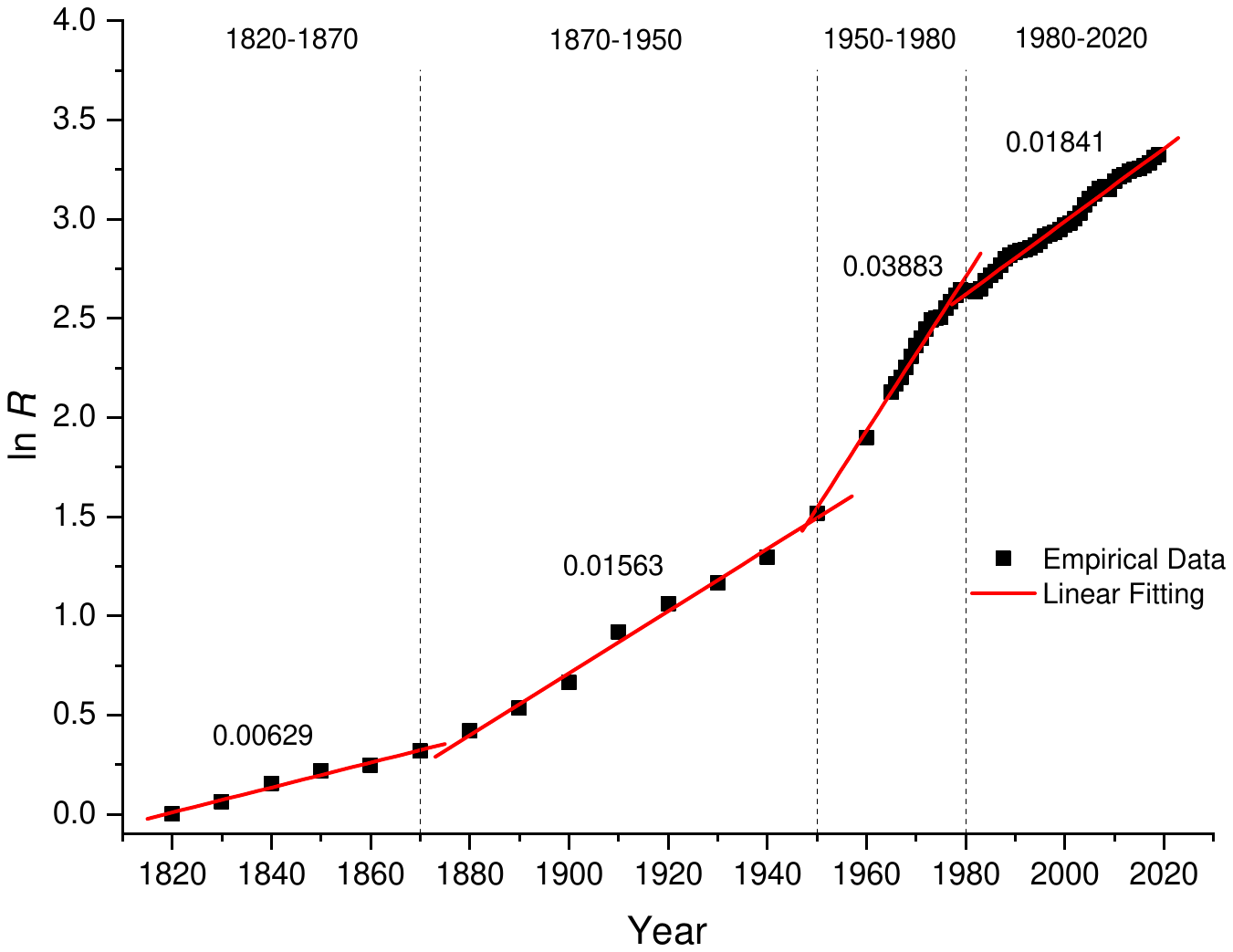}
\end{center}
\vspace*{-8mm}
\caption{Data Source: Ref. \cite{energy} is the source of data for energy consumption $R$, given as a percentage of its original value in 1820.}
\label{fig2}
\end{figure}

\begin{figure}
[!htb]
\begin{center}
\hspace*{-10mm}\includegraphics[width= 13.7cm]{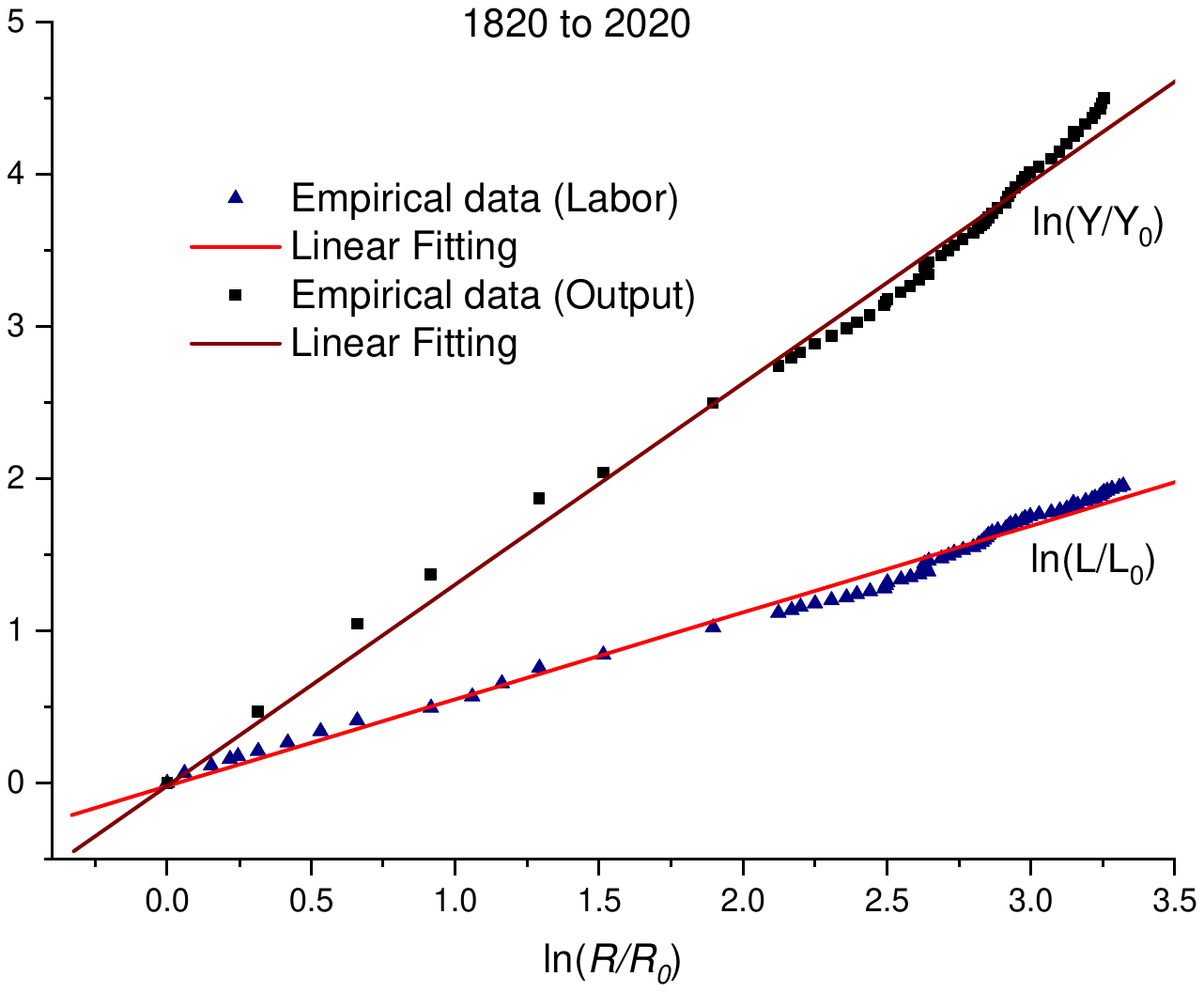}
\end{center}
\vspace*{-8mm}
\caption{The population $L$ data comes from Ref. \cite{population}, GNP $Y$ from Ref. \cite{gdp}, and energy consumption $R$ from Ref. \cite{energy}. $L_0$, $Y_0$ and $R_0$ are the variables’ respective values in the year 1820.}
\label{fig3}
\end{figure}

Figure \ref{fig2} clearly shows that the variable $R$ exhibits different growth rates across roughly four time periods: 1820-1870, 1870-1950, 1950-1980, and 1980-2020. The growth rate $\lambda_R$ of the variable $R$ is approximately constant for each of these periods. As a result, the asymptotic exponential solution for $R$ in equation (\ref{rlc_solow}) appears to be well adjusted to explain the temporal evolution of the empirical quantity chosen to represent this variable in each of these periods.  
Figure \ref{fig3} depicts the linear fit for the relationship between the variables $\ln(Y)$ and $\ln(N)$ with the variable $\ln(R)$ for the entire period 1820 to 2020.

We propose a “quasi-equilibrium hypothesis” to explain the results shown in Figures \ref{fig2} and \ref{fig3} in terms of equations (\ref{new2}) and (\ref{new9}). It is assumed that $X_1$ and $X_2$ spend the majority of their time, from 1820 to 2020, near some fixed point $X_1^{**}$ and $X_2^{**}$ (\ref{new3}).
This quasi-equilibrium hypothesis suggests that the dynamics of the original variables $K$, $N$ and $R$ can be approximated by the differential equations in (\ref{new9}) with $\lambda_K$,
$\lambda_N$ and $\lambda_R$ given in (\ref{new8a}). Thus, the differential relationships between the variables $Y$ and $N$ with the variable $R$ are given as follows:
\begin{equation}
\label{difKLcomR}
\frac{dY}{Y}=\beta^{-1}\frac{dR}{R},\;\;\frac{dN}{N}=\left(\frac{1-\alpha_1}{\beta\alpha_2}\right)\frac{dR}{R},
\end{equation}
where equation (\ref{A34}) gives the differential equation for $Y$. The following allometric relations are implied by the differential relations in (\ref{difKLcomR}):
\begin{equation}
\label{ajuste3}
\frac{Y}{Y_0}=\left(\frac{R}{R_0}\right)^{\beta^{-1}},\;\;\frac{N}{N_0}=\left(\frac{R}{R_0}\right)^{\frac{1-\alpha_1}{\beta\alpha_2}}.
\end{equation}
We obtain an estimate for the exponents that appear in the allometric relationships (\ref{ajuste3}) by performing a nonlinear fit on the empirical data shown in Figure \ref{fig1}:
\begin{equation}
\label{ajuste2}
\beta^{-1}=1.34037\;\Rightarrow\beta=0.74606;\;\;\;\frac{1-\alpha_1}{\beta\alpha_2}=0.58282.
\end{equation}

\begin{figure}
[!htb]
\begin{center}
\hspace*{-10mm}\includegraphics[width= 13.7cm]{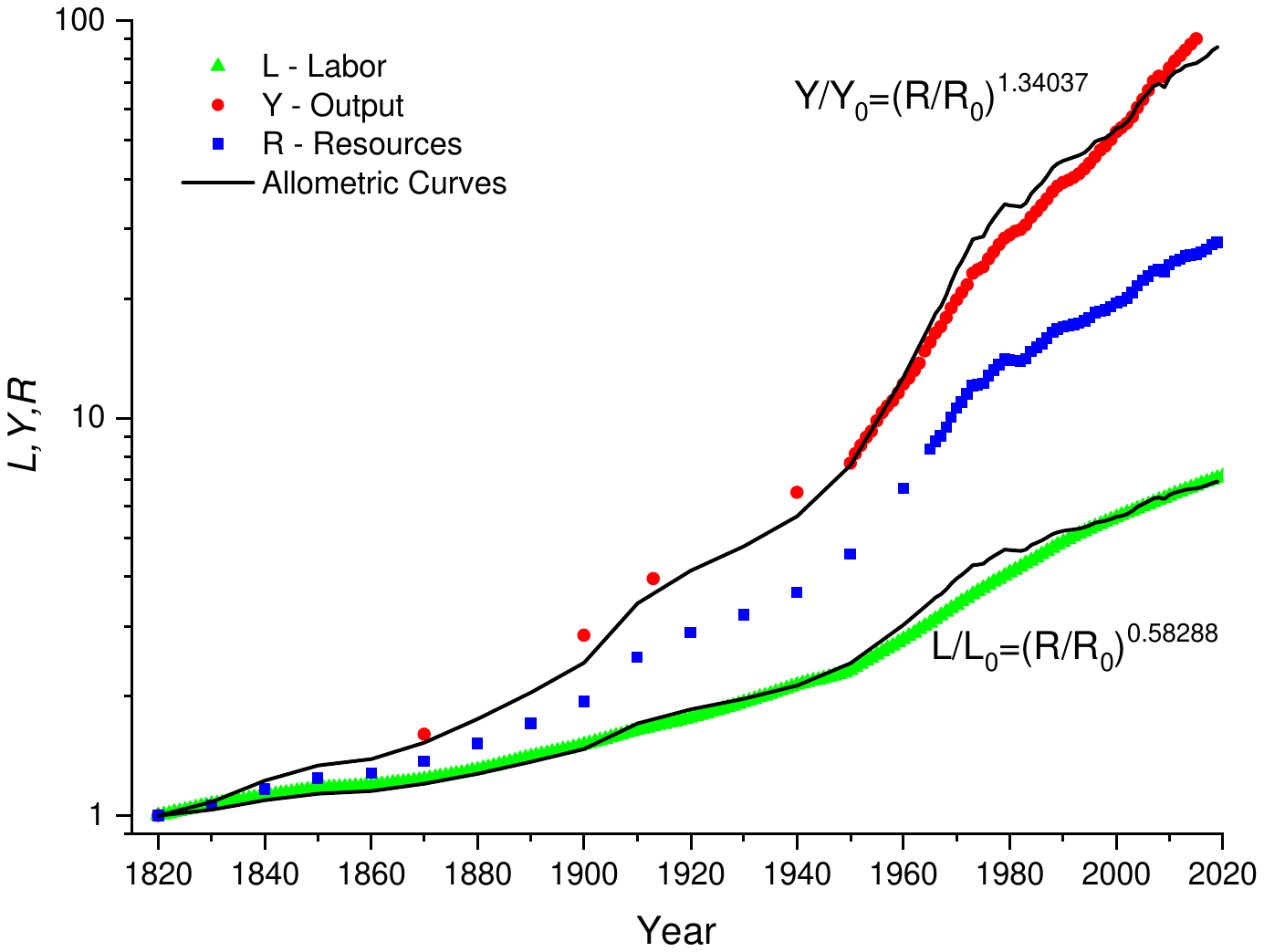}
\end{center}
\vspace*{-8mm}
\caption{ The population $N$ data comes from Ref. \cite{population}, GNP $Y$ from Ref. \cite{gdp}, and energy consumption $R$ from Ref. \cite{energy}. $N_0$, $Y_0$ and $R_0$ are the variables respective values in the year 1820. Nonlinear fitting of the allometric functions in (\ref{ajuste3}) yields the exponents of the allometric curves.}
\label{fig4}
\end{figure}

In Figure \ref{fig4}, we compare the empirical time evolution of the variables $Y$ and $N$ with the allometric adjustment of the empirical variable $R$, using the relationships in  (\ref{ajuste3}) with the exponents obtained in (\ref{ajuste2}). 
Based on the findings, we conclude that the quasi-equilibrium hypothesis and the exponents defined in (\ref{ajuste2}) are correct. 

To summarize, equations (\ref{new2}) and (\ref{new9}), as well as their quasi-equilibrium solutions in (\ref{rlc_solow}) are functions of the model’s various parameters (see equation (\ref{A4})). These parameters must vary between 1820 and 2020, implying different growth rates $\lambda_R$, as shown in Figure \ref{fig2}. Concurrently, these fluctuations must imply a change in the fixed point (\ref{new3}), but the relationships given in (\ref{ajuste2}) remain constant over the entire period under consideration. Despite these fluctuations, the model’s solution is always close to an equilibrium solution corresponding to a given fixed point, implying that the variables evolution should roughly obey the relations given in (\ref{ajuste3}).

We can rewrite the expressions for $\lambda_R$, $\lambda_N$ and $\lambda_K$ (respectively $\lambda_Y$) in (\ref{new8a}) as
\begin{equation}
\label{lambdaR}
\lambda_R=\frac{{b}{\bar k}}{1+{\bar k}\left(\displaystyle\frac{1-\alpha_1}{\beta\alpha_2}\right)},\;\;\lambda_N=\frac{1-\alpha_1}{\beta\alpha_2}\lambda_R,\;\;
\lambda_Y=\beta^{-1}\lambda_R.
\end{equation}
Substituting the estimated values in Eq. (\ref{ajuste2}) into Eq. (\ref{lambdaR}), we obtain:
\begin{equation}
\label{lambdaR1}
\lambda_R=\frac{{b}{\bar k}}{1+0.58282\,{\bar k}},\;\;\lambda_N=0.58282\,\lambda_R,\;\;\lambda_Y=1.34037\,\lambda_R.
\end{equation}
As a result, we can conclude that fluctuations in the growth rates of the variables $N$, $K$, and $R$ for the entire period 1820-2020 are caused by variations in the parameter $b$ and/or the parameter ${\bar k}$ , which is defined in equation (\ref{parametros_bk}) as a function of the parameters $g_0$, $e_0$ and $r_0$.

The main component of the interpretation is the global stability property for the fixed point (\ref{new3}), which states that for any initial condition, the corresponding solution converges asymptotically to the model’s fixed point. This fact implies two non-exclusive possibilities: the change from an old fixed point to a new one is very small, and thus the solution will be close to this new fixed point, or the convergence time to the new fixed point is very small when compared to the time interval of 200 years associated with the entire period.

We calculate the following value for the relationship given in (\ref{A34}) based on the values obtained in (\ref{ajuste2}): 
\begin{equation}
\label{A33_1820_2020}
\frac{\lambda_K}{\lambda_N}=\frac{\lambda_Y}{\lambda_N}=\frac{\alpha_2}{1-\alpha_1}=\frac{1.34037}{0.58282}=2.29980>1,
\end{equation}
which implies $\alpha_1+\alpha_2>1$, as shown in equation (\ref{A34a}). As a result, we can see that this result is consistent with what we call a non-Malthusian model, in which the per capita relationship between the variables $Y$ and $N$ is always increasing. In other words, per capita wealth is continually rising over the time period under consideration. 

Taking into account the relations in Eq. (\ref{cond_expoentes}), we obtain:
\begin{equation}
\label{delta2}
\alpha_1+\alpha_2=\beta^{-1}\;\Rightarrow\; \alpha_1+\alpha_2=1.34037.
\end{equation}
Solving the system of equations (\ref{A33_1820_2020}) and (\ref{delta2}) with respect to $\alpha_1$ and $\alpha_2$ we obtain:
\begin{equation}
\label{alpha_estima}
\alpha_1=0.73814,\;\;\alpha_2=0.60223.
\end{equation}

\subsection{Period of 0-1820}

\begin{figure}
[!htb]
\begin{center}
\hspace*{-10mm}\includegraphics[width= 13.7cm]{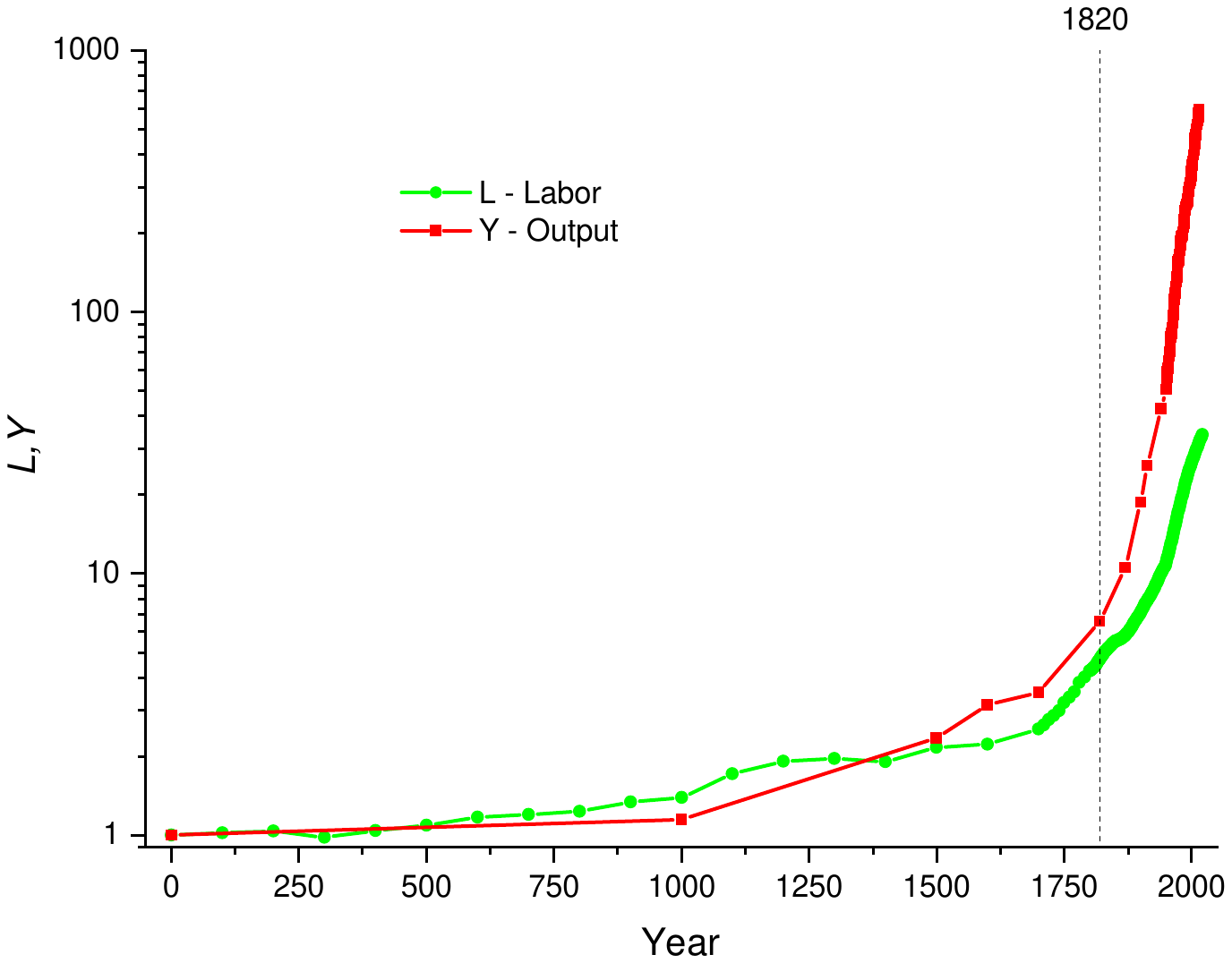}
\end{center}
\vspace*{-8mm}
\caption{Ref. \cite{population} contains the data for population $N$, and GNP is found in Ref. \cite{gdp}. Both variables are given as a percentage of their initial reference time at 0. A monolog scale is used for the vertical axis.}
\label{fig5}
\end{figure}

We do not have data for the resource variable $R$ for the period 0-1820, so we will only look at the time evolution of variables
$N$ and $Y$. Figure \ref{fig5} depicts these variables for the entire period 0-2020, including the period 1820-2020, which was examined in the previous subsection, for comparison purposes.

\begin{figure}
[!htb]
\begin{center}
\hspace*{-10mm}\includegraphics[width= 13.7cm]{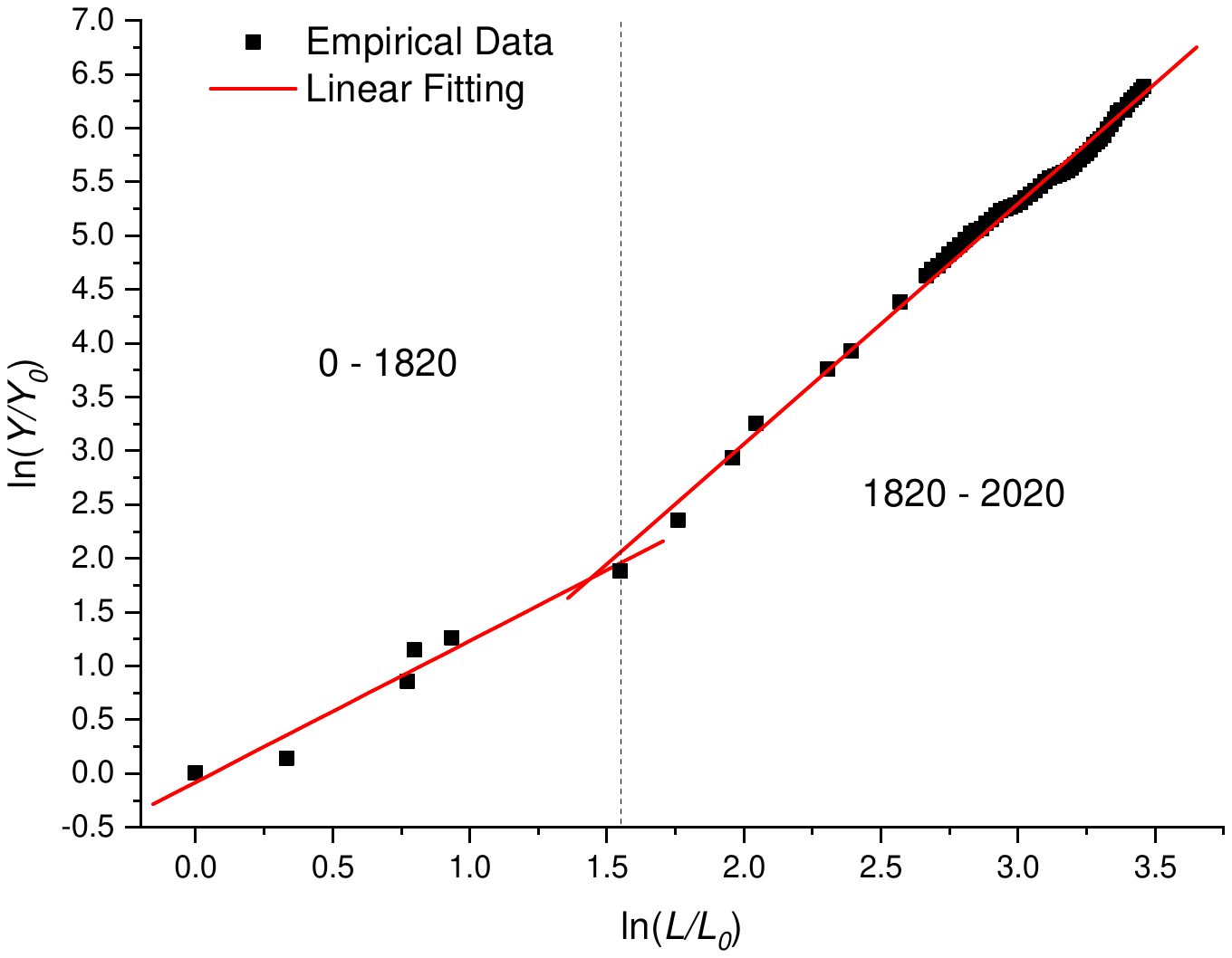}
\end{center}
\vspace*{-8mm}
\caption{Ref. \cite{population} contains the data for population $N$, and GNP is found in Ref. \cite{gdp}. Both variables are given as a percentage of their initial reference time at 0.}
\label{fig6}
\end{figure}

\begin{figure}
[!htb]
\begin{center}
\hspace*{-10mm}\includegraphics[width= 13.7cm]{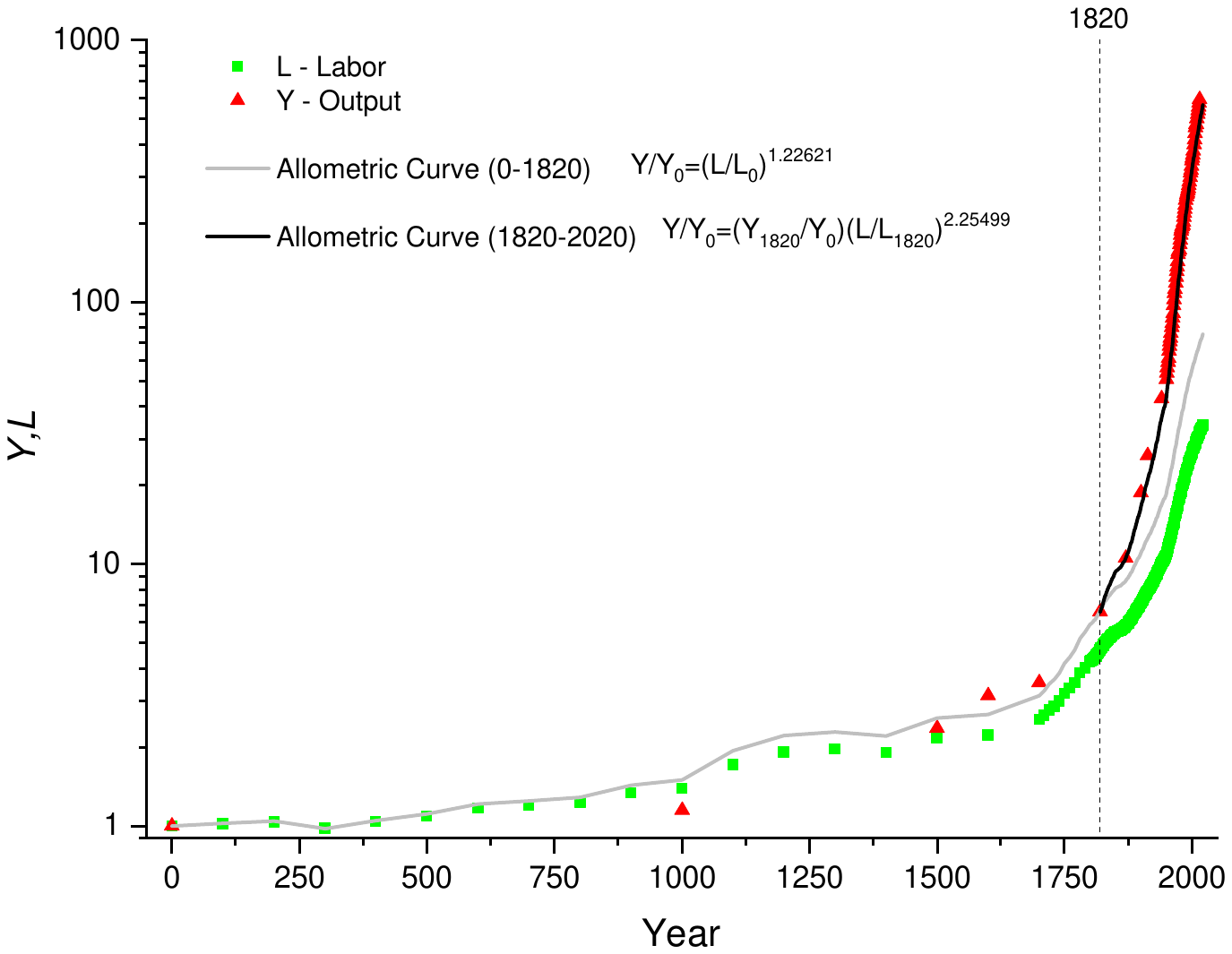}
\end{center}
\vspace*{-8mm}
\caption{Ref. \cite{population} contains the data for population $L$, and GNP is found in Ref. \cite{gdp}. $L_0$ and $Y_0$ are the respective values of the variables in the year 0. $L_{1820}$ and $Y_{1820}$ are the respective values of the variables in the year 1820.The exponents of the allometrics curves are obtained by non linear fitting of the allometric function in (\ref{alometrica_YL}).}
\label{fig7}
\end{figure}

Assuming the quasi-equilibrium hypothesis, we can derive the following allometric relationship between variables $N$ and $Y$ from equations (\ref{new9}) and (\ref{new8a}):
\begin{equation}
\label{alometrica_YL}
\frac{dY}{Y}=\frac{\alpha_2}{1-\alpha_1}\frac{dN}{N}\;\Rightarrow\; \frac{Y}{Y_0}=\left(\frac{N}{N_0}\right)^{\frac{\alpha_2}{1-\alpha_1}}.
\end{equation}
The figure \ref{fig6} shows respectively the linear fit for the relationship between the variables $\ln(Y)$ and $\ln(N)$ for the two periods under consideration: 0-1820 and 1820-2020. 
Making a nonlinear fit to obtain the exponent of the allometric relation in (\ref{alometrica_YL}) we obtain:
\begin{equation}
\label{expalometricoYL0a2020}
\frac{\alpha_2}{1-\alpha_1}=\left\{
\begin{array}{l}
1.22937\;(0-1820) \\ \\
2.32991\;(1820-2020)
\end{array}\right.
\end{equation}
Finally, in Figure \ref{fig7}, we compare variable $Y$’s temporal evolution to variable $N$’s allometric scaling.

Figures \ref{fig6} and \ref{fig7} demonstrate that the quasi-equilibrium hypothesis fits the empirical data. There is a clear transition in the exponent $\alpha_2/(1-\alpha_1)$ between the two periods under consideration.  However, we must be cautious in claiming that the exponent of the allometric relation
(\ref{expalometricoYL0a2020}) has a unique value for the period 0-1820, because, between the years 0 and 1500, we only have a single datum for the value of the variable Y, corresponding to the year 1000 (Figure \ref{fig5}), and which clearly looks well below the rescaled allometric curve shown in Figure \ref{fig7}.

Figure \ref{fig7} suggests that the periods 1500-1820 and 1820-2020 are distinguished by two distinct allometric relationships, with well-defined exponents for each period. This exponent has undergone a transition, which must have occurred around the year 1820. On the other hand, it is possible to speculate that the value of the exponent $\alpha_2/(1-\alpha_1)$ in the period 0-1000 was less than the value calculated in (\ref{expalometricoYL0a2020}) and possibly less than $1$. This would imply a Malthusian evolution period with $\alpha_1+\alpha_2<1$.We can observe this fact by measuring, separately, the exponents of the allometric relation (\ref{alometrica_YL}) for the respective periods of $0-1000AC$ and $1000-1820AC$:
\begin{equation}
\label{expalometricoYL0a1820}
\frac{\alpha_2}{1-\alpha_1}=\left\{
\begin{array}{l}
0.42132\;(0-1000) \\ \\
1.46614\;(1000-1820)
\end{array}\right.
\end{equation}

It is worth noting that there are periods in the time interval 0-1500 when the variable $N$ decreases, and by the quasi-equilibrium hypothesis and the expressions in (\ref{new8a}), the variable Y should also decrease during this period and ${\bar k}<0$. Unfortunately, we do not have enough empirical values of the variable Y for this period, so testing the validity of this property associated with the quasi-equilibrium hypothesis is impossible.
If the variable $Y$ does not grow in the same direction as the variable $N$, we must abandon the quasi-equilibrium hypothesis and try to explain the dynamics of the variables under consideration using non-equilibrium solutions.

One final observation is that the values of exponent $\alpha_2/(1-\alpha_1)$ for the period 1820-2020 were obtained in two ways: indirectly from equation (\ref{A33_1820_2020}) and directly from a nonlinear fit in equation (\ref{expalometricoYL0a2020}). We can see that these values are very close, indicating that the quasi-equilibrium hypothesis appears to be a good hypothesis for explaining the empirical time evolution of the variables under consideration.

\section{Exponential growth rate of carrying capacity limit}
\label{exponential}

What we showed in the previous section is that a production process, modeled by the equations in (\ref{int5}) and with $R_T=\infty$, necessarily implies exponentially growing asymptotic solutions for the variables $K$, $N$ and $R$, which seems to explain the empirical temporal evolution of these variables. This type of solution can only appear if we make the non-trivial assumption that there are no constraints or limitations on the $R_T$ carrying capacity limit. In this section we will try to assess under what conditions this hypothesis can be considered reasonable.

The idea here is to add to the model described by Eq. (\ref{int5}) an exponential growth equation for the carrying capacity limit $R_T$:
\begin{equation}
	\label{linRT}
	\frac{dR_T}{dt}=\mu R_T,
\end{equation}
where the exponential growth rate  $\mu>0$ for the carrying capacity limit
may be achieved, for instance, by technological improvements or a better allocation of goods. 

Using the temporal reparametrization $t'\rightarrow t/d$, we obtain a system of ODE's equal to that of Eq.~(\ref{A4}) plus the re-parametrized equation for $R_T$. It is worth noting that the parameter $R_T$ is now a variable of the obtained four-dimensional ODE's system. This system, when written in its QP form in (\ref{qp1}), will imply the following three quasi-monomial as defined in (\ref{defx}):
\begin{equation}
\label{monRT} U_1=C^{\alpha_1-1}L^{\alpha_2},\;\;U_2=\frac{C^{\beta_1} N^{\beta_2}}{R},\;\;U_3=\frac{C^{\beta_1} N^{\beta_2}}{R_T},
\end{equation}
which must satisfy a three-dimensional Lotka-Volterra system as in (\ref{LVdef}).
Now, similarly to what was done in (\ref{monomiosX}), we define the following rescaled monomial variables:
\begin{equation}
\label{resRT} X_1=\frac{s}{d}U_1,\;\; X_2=\frac{r}{d}U_2,\;\;X_3=\frac{g}{d}U_3.
\end{equation}

The variables defined above will obey the following Lotka-Volterra system:
\begin{eqnarray}
\label{lvRT}
	\frac{dX_1}{dt} & = &  X_1\left(1-\alpha_1+\displaystyle{\bar{b}}\alpha_2-(1-\alpha_1)X_1-\alpha_2X_2\right),
	\nonumber\\
	\frac{dX_2}{dt} & = &  X_2\left(-\beta_1+\displaystyle{\bar{b}}\beta_2+\beta_1X_1-(\bar{k}+\beta_2)X_2+X_3\right),
	\\
	\frac{dX_3}{dt} & = &  X_3\left(-\beta_1+\displaystyle{\bar{b}}\beta_2-\bar{\mu}+\beta_1X_1-\beta_2X_2\right),\nonumber
\end{eqnarray}
with $\bar{\mu}=\mu/d$ being the time re-parametrized exponential growth rate for the carrying capacity limit $R_T$. Eq.(\ref{lvRT}) differs from Eq.~(\ref{eq2}) only by
the linear part of the corresponding equation for the variable $X_3$. Indeed, when $\bar{\mu}=0$  Eq.~(\ref{lvRT}) becomes equal to Eq.~({\ref{eq2}).  
Finally, the growth rates for the variables $K$, $N$ and $R$ will be given by the same expressions (\ref{new0}) obtained for the case where $\bar{\mu}=0$.

The system of equations (\ref{lvRT}) has an interior fixed point whose components are given by
\begin{equation}
\label{newA23}
X_1^*=1+\bar\mu\frac{\alpha_2}{\beta\alpha_2},\;\;X_2^*=\displaystyle{\bar b}-\bar\mu\frac{1-\alpha_1}{\beta\alpha_2},\;\;X_3^*=\displaystyle{\bar b}{\bar k}-\bar\mu\left(\frac{\beta\alpha_2+(1-\alpha_1){\bar k}}{\beta\alpha_2}\right).
\end{equation}
Considering ${\bar b}>0$, $\bar k>0$ and $\alpha_1\leq 1$, we can demonstrate that $X^*_2>0$ and $X^*_3>0$ if the respective inequalities are satisfied:
\begin{equation}
{\bar \mu}<{\bar\mu}_1=\frac{\bar b \beta\alpha_2}{1-\alpha_1},\;\;\bar\mu<{\bar\mu}_2=\frac{\bar k {\bar b} \beta\alpha_2}{\beta\alpha_2+(1-\alpha_1)\bar k}.
\end{equation} 
Furthermore, it is easy to show by direct verification that
\begin{equation}
{\bar\mu}_2-{\bar\mu}_1=-\frac{{\bar b}(\beta\alpha_2)^2}{(1-\alpha_1)({\bar k}(1-\alpha_1)+\beta\alpha_2)}<0.
\end{equation}
Thus, the interior fixed point (\ref{newA23}) only exists in the positive orthant ${\R}^3_+$ if the following condition is satisfied:
\begin{equation}
\label{pfcond} 0\leq\bar\mu<{\bar \mu}_2=\frac{\bar k \bar b \beta\alpha_2}{\beta\alpha_2+(1-\alpha_1)\bar k}.
\end{equation}
Now, substituting the expressions of this fixed point into the growth rates in (\ref{new0}), we get:
\begin{equation}
\label{taxasPFI}\lambda_K=\frac{\bar \mu \alpha_2}{\beta\alpha_2},\;\;\lambda_N=\frac{\bar\mu (1-\alpha_1)}{\beta\alpha_2},\;\;\lambda_R=\bar \mu.
\end{equation} 

As a result, the fixed point (\ref{newA23}) represents a stationary solution to the system (\ref{lvRT}) with exponential growth for the variables $K$, $N$, and $R$. It is worth noting in particular that the growth rate of carrying capacity $\lambda_R$ is equal to the growth rate of carrying capacity limit $R_T$. Furthermore, for $\mu=0$ the fixed point (\ref{newA23}) becomes the fixed point defined in (\ref{A23}).

Another fixed point of the system (\ref{lvRT}) is given by
\begin{equation}
\label{newnew3}
X_1^{**}=\frac{\beta\alpha_2+{\bar k}\Delta_3}{\beta\alpha_2+{\bar k}(1-\alpha_1)},\;\;X_2^{**}=
\frac{{\bar b}\beta\alpha_2}{\beta\alpha_2+{\bar k}(1-\alpha_1)},\;\;X_3^{**}=0.
\end{equation}

This fixed point has non-zero components that are the same as the fixed point in (\ref{new3}). It is an invariant face such that $X_3=0$, which corresponds to a two-dimensional Lotka-Volterra system producing $X_3=0$ in Eq (\ref{lvRT}). This system is equivalent to Eq.  
(\ref{new2}). As shown in Section 4, any solution with positive components $X_1$ and $X_2$ contained in this face will converge to the fixed point (\ref{newnew3}). In this case, however, unlike in (\ref{new2}), the fixed point (\ref{newnew3}) is a stationary solution to the three-dimensional system (\ref{lvRT}). As a result, we investigate the stability of this fixed point as a solution to the system (\ref{lvRT}).

As the fixed point (\ref{newnew3}) is a global attractor for any solution that is contained in the face $X_3=0$ and with components $X_1,X_2>0$, then two stable manifolds of this fixed point must be contained in this face and the respective eigenvalues, denoted by $\lambda_1^{**}$ and $\lambda_2^{**}$, will certainly have a negative real part. The third eigenvalue, denoted by $\lambda_3^{**}$, will  correspond to a linear manifold containing the fixed point and transverse to the face at this point. Its value can easily be calculated and is given by:
\begin{equation}
\label{pfcond1}
\lambda_3^{**}=\bar\mu_2-\bar\mu,
\end{equation}
where $\bar\mu_2$ is defined in (\ref{pfcond}).

When $\bar\mu>\bar\mu_2$, equation (\ref{pfcond1}) implies that $\lambda_3^{**}<0$, and the fixed point (\ref{newnew3}) is a stable attractor. Furthermore, according to equation (\ref{pfcond}), the fixed point (\ref{newA23}) will not exist in the positive orthant ${\R}^3_+$. As a result, the fixed point (\ref{newnew3}) represents an asymptotic stable solution to Eq. (\ref{lvRT}), and the respective growth rates of $K$, $N$ and $R$ are given by the expressions in (\ref{new8a}), which correspond to a stationary solution to Eq. ( \ref{new2}) modeling a system with infinite carrying capacity limit ($R_T=\infty$).

As a result, we demonstrate that we do not need to assume that $R_T=\infty$, but rather that $R_T$ has an exponential growth rate given by (\ref{linRT}) and that ${\bar\mu}>{\bar\mu}_2$.
This last inequality, considered in the expression of
$\lambda_R$ in (\ref{new8a}), will imply
\begin{equation}
\label{condinf}
\mu>\lambda_R=\frac{ b \bar k  \beta\alpha_2}{\beta\alpha_2+(1-\alpha_1)\bar k}.
\end{equation} 
This clearly shows that the growth of carrying capacity will always be smaller than the growth of support capacity for the stable attractor solution represented by the fixed point (\ref{newnew3}), that is,
\begin{equation}
\label{condinf1}
\frac{\dot R}{R}<\frac{{\dot R}_T}{R_T}.
\end{equation}

Another way to look at this question is to say that if the carrying capacity limit grows exponentially as shown in (\ref{linRT}), and condition (\ref{condinf}) is met, then the model described by Eq. (\ref{new2}) with the hypothesis of $R_T=\infty$ is a good approximation to explain the exponential growth of the variables $K$, $N$ and $R$.

To test the validity of equation (\ref{condinf1}) for the data examined in the previous section, compare the increase in variable R, which represents the amount of equivalent energy consumed, with the increase in variable $R$, which must represent the reserves of available sources of consumed energy.

In the left panel of Figure \ref{fig10}, we show the time evolution of the equivalent energy consumed from 1820 to 2020 
(as in Figure \ref{fig1}), but this time we identify the main sources of consumed energy. We can see that oil (petroleum), natural gas, and coal always account for more than $80\%$ of total energy consumption after 1940. Unfortunately, we do not have information on the availability of gas and coal reserves for the time period under consideration.

\begin{figure}
[!htb]
\begin{center}
\hspace*{-5mm}\includegraphics[width= 6.7cm]{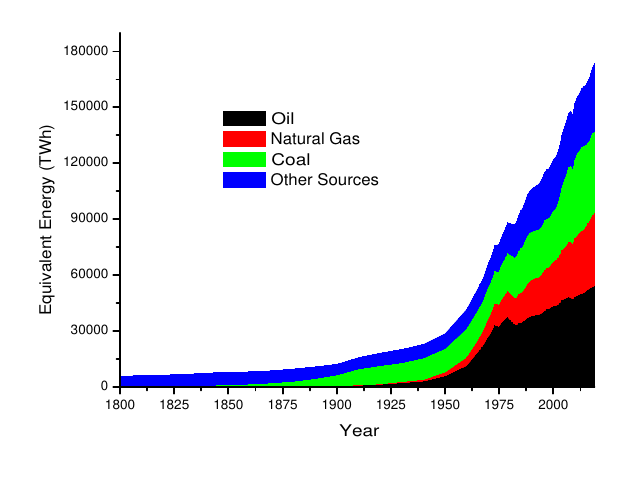}
\hspace*{-9mm}\includegraphics[width= 6.7cm]{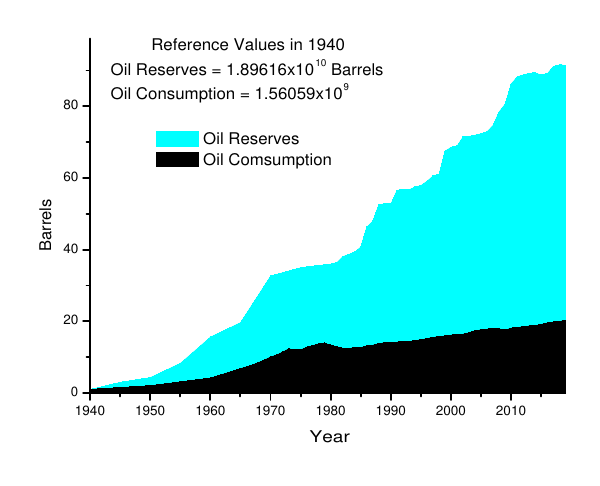}
\end{center}
\vspace*{-8mm}
\caption{For equivalent energy data, see Ref. \cite{energy}, and for oil reserves, see Ref. \cite{reserves}. The variables in the right panel are presented as a percentage of their 1940 values.}
\label{fig10}
\end{figure}

On the other hand, we have relatively accurate data for the evolution of oil reserves since 1940, which allows us to compare it to the evolution of its consumption. Figure 10’s right panel depicts this. We can clearly see that the relationship given in equation (\ref{condinf1}) is satisfied, supporting the idea that carrying capacity limit growth rate  $R_T$ exceeds the variable $R$ growth rate. Naturally, we extrapolate the behavior observed for oil to other energy sources, as well as to the entire period considered from 1820 to 2020.

\section{Concluding Remarks}
\label{fim}

The main property of the model represented by equations (\ref{new2}) and (\ref{new9}) is the fact that the fixed point (stationary solutions) of the system (\ref{new2}) implies exponential growth rates for the model variables.
It is very important to point out that different values of the parameters that determine the ODE system in (\ref{new2}) imply different fixed point (\ref{new3}), leading to different exponential growth rates in (\ref{new8a}). However, the relationships between these growth rates imply the allometric relationships defined in equations (\ref{ajuste3}) and (\ref{alometrica_YL}), which depend solely on the scale exponents of the production and support functions defined respectively in the equation (\ref{int6}), that is, the exponents $\alpha_1$, $\alpha_2$ and $\beta$.

The main hypothesis of this work was to state that the successive periods of exponential growth of the human population can be described by the exponential solution (\ref{new9}). This hypothesis was called the quasi-equilibrium hypothesis and
it is based on two assertions: (i) the variables $X_1$ and $X_2$ of the system (\ref{new2}) are almost always close to the fixed point (\ref{new9}); (ii) the exponential growth rates of variables can change over time.

The empirical analysis carried out in section 5 corroborates statement (i) since we found that the model variables approximately satisfy the allometric relations defined in equations (\ref{ajuste3}) and (\ref{alometrica_YL}). This fact allowed to obtain empirical estimates for the exponents of these relations, according to Eqs. (\ref{ajuste2}), (\ref{A33_1820_2020}), (\ref{expalometricoYL0a2020}) and (\ref{expalometricoYL0a1820}). The main property of the system (\ref{new2}) that supports this hypothesis is the global stability associated with the fixed point of this system.
Evidence for statement (ii) comes from the observation that the exponential growth rates for the model variables tend to have different values for different time intervals (as can be seen in figure \ref{fig8} and also in figure \ref{fig1}).

The allometric relationships obtained empirically in section 5 seem to indicate that the scale exponents of the production and support functions must be associated with long-term structures, because despite the variation in the growth rates of the variables, the allometric relationships between them remain the same for long periods.

Once the scale exponents of the support and production functions are fixed, the growth rates of the model variables in (\ref{lambdaR}) will only depend on the parameters $b$ and ${\bar k}$. This fact leads us to the conclusion that variations in growth rates, for periods of time smaller than the time in which the scale exponents remain constant, must be a consequence of variations in those parameters.
It is also worth noting that once the scale exponents are fixed, the system (\ref{new2}) only depends on the parameters $\bar b$ and $\bar k$. Thus, the dynamics of return to equilibrium will be determined only by these parameters.

The quasi-permanent growth of the human population over the last $12,000$ years can be explained, in the light of our model, as being a consequence of the fact that the parameter ${\bar k}$ must be positive most of the time, which will imply permanent exponential population growth. If we look at the definition of this parameter in (\ref{parametros_bk}) we will have
$${\bar k}>0\;\;\Rightarrow\;\;g_0>e_0.$$
That is, the permanent growth of the human population is associated with a permanent capacity to create more than destroy carrying capacity.
Taking into account the analysis made in section 6, this property of increasing carrying capacity is associated with a permanent increase in its growth limits, which is modeled by equation (\ref{linRT}), where we have a permanent growth of the carrying capacity limit $R_T$. This phenomenon was illustrated in figure \ref{fig10}.

An interesting question is whether this process can continue indefinitely. In fact, observation of the growth process over the last 12,000 years,  and in particular its spectacular acceleration over the last $250$ years (see figures \ref{fig8} and \ref{fig7}), does not seem to suggest a saturation process. However, it is interesting to analyze what would happen if variable $R_T$ saturate, that is, the existence of a finite limit for the growth of the carrying capacity. In this case, the model to describe the  economic and population growth would be given by the system of equations (\ref{eq2}) and (\ref{new0}), implying asymptotically in a non-growth economy (as can be seen in equations (\ref{pfklr}) and (\ref{estagnante}).

Another interesting point to discuss is the consequence of having hypothesized that the support function has constant returns to scale and is related to the production function based on a scale relationship defined by the exponent $\beta$, as defined in the equation 
(\ref{int6}) and (\ref{cond_expoentes}), which allow us to write:
$$Y=S^{\beta^{-1}}=S^{\alpha_1+\alpha_2}.$$
The relation above allows us to interpret the return-to-scale of the production function in terms of the exponent $\beta^{-1}$. In terms of the quasi-equilibrium hypothesis, this relation allows interpreting the difference between Malthusian versus non-Malthusian growth as being determined by the value of the exponent $\beta^{-1}$, with the transition between the two types of growth determined by the Solow model ($\beta^{-1}=1$). 

It is quite clear from our empirical analysis in Section 5 that the production function does not have constant return-to-scale and therefore the Solow model is not suitable for describing economic growth over long time scales.
Furthermore, when we observe the per capita ratios in (\ref{rlc_percapita}), we see that they do not depend only on the ratio $s_0/d$, as in the case of the Solow model, implying a greater range of possible asymptotic states  that  also depend on the parameters $b$, $r$, ${\bar k}$ and, naturally, on the scale exponents of the production and support functions.

\appendix

\section{The Global Stability Properties}

The purpose of this appendix is to demonstrate that the fixed points at (\ref{A23}) and (\ref{new3}) are globally stable, that is, they are attractors of all solutions defined in the positive octant ${\R}^ 3_+$ and ${\R}^2_+$ of the respective ODE's systems given in (\ref{eq2}) and (\ref{new2}).  We will use the results obtained in the reference \cite{redheffer}, where a theory about the global stability of interior fixed points in Lotka-Volterra type systems is developed. This theory is based on the construction of Lyapunov functions defined from admissible matrices and on the development of a graph theory that allows characterizing different sufficient conditions of global stability for the interior fixed points.
It is not our objective here to develop all aspects of this theory in the analysis of our problem, but only to obtain the stability conditions that are sufficient to justify the results and applications made in this work.

\subsection{Global Stability of The Fixed Point in (\ref{A23})}

The function $f=X_3/X_2-{\bar k}$ is a semi-invariant of the system of ODE's given in (\ref{eq2}), since it obeys the following differential equation:
\begin{equation}
\label{app1}
\frac{df}{dt}=\frac{d}{dt}\left(\frac{X_3}{X_2}\right)=\frac{{\dot X}_3X_2-X_3{\dot X}_2}{X_2^2}=\frac{F_3X_2-X_3F_2}{X_2^2}=-Y_3\, f.
\end{equation}
The differential equation (\ref{app1}) implies that $f=0$ is an invariant surface of the dynamics of the solutions of the system (\ref{eq2}), that is, any initial condition defined at a given time $t_0=0$ and belonging to this surface will imply that the corresponding solution will be contained in it. It is worth noting that the invariant surface $f=0$ is nothing more than the plane defined by the relation $X_3={\bar k}X_2$.

It is easy to show, from the equation (\ref{app1}), that any solution, defined in its maximum range $[0,t_{max})$ and contained in the positive octant ${\R}^3_+$, will converge asymptotically to the invariant surface $f=0$ when $t\rightarrow t_{max}$. In other words, we have the following asymptotic relationship:
\begin{equation}
\label{app2a}
X_3(t)={\bar k}X_2(t)\;\;(t\rightarrow t_{max})
\end{equation}
Substituting the relation (\ref{app2a}) into equation (\ref{eq2}) we obtain the following two-dimensional Lotka-Volterra system:
\begin{eqnarray}
\label{app2b}
\frac{dX_1}{dt}&=&X_1\left(1-\alpha_1+\bar{b}\alpha_2-(1-\alpha_1)X_1-\alpha_2X_2\right),\nonumber\\
\frac{dX_2}{dt}&=&X_2\left(-\beta_1+\bar{b}\beta_2+\beta_1X_1-\beta_2X_2\right).
\end{eqnarray}
This system determines the dynamics of all solutions contained in the invariant surface $f=0$. In this way, all solutions contained in the positive orthant will be asymptotically determined by the ODE's system (\ref{app2b}).

The fixed point (\ref{A23}) is contained in the invariant plane $f=0$ and its first two components determine the fixed point of the system (\ref{app2b}), that is, this fixed point will be given by:
\begin{equation}
\label{app3}
X_1^*=1,\;\;X_2^*={\bar b}.
\end{equation}

Using the construction of Lyapunov functions for Lotka-Volterra systems defined in the reference \cite{redheffer}, we define the following function:
\begin{equation}
\label{app4}
V=X_1^*\left[\left(\frac{X_1}{X_1^*}\right)-\ln \left(\frac{X_1}{X_1^*}\right)-1\right]
+\frac{\alpha_2}{\beta_1}X_2^*\left[\left(\frac{X_2}{X_2^*}\right)-\ln \left(\frac{X_2}{X_2^*}\right)-1\right],
\end{equation}
whose derivative with respect to time $t$ will be given by
\begin{equation}
\label{app5}
\frac{dV}{dt}=-\left[\left(1-\alpha_1\right)\left(X_1-X_1^{*}\right)^2+\frac{\alpha_2\beta_2}{\beta_1}\left(X_2-X_2^{*}\right)^2\right]\leq 0
\end{equation}

The equations (\ref{app4}) and (\ref{app5}) imply that any solution, contained in the invariant plane $f=0$, defined in its maximum interval $[0,t_{max})$ and with condition initial such that
$X_1(0)>0$ and $X_2(0)>0$, will be limited and permanent (see reference \cite{redheffer}), so they will be defined for all time $t>0$, that is, we will have $t_{max}=\infty$.
Using the LaSalle principle \cite{lasalle} for the Lyapunov function (\ref{app4}), we can conclude that these solutions will converge asymptotically ($t\rightarrow\infty$) to the region where
$dV/dt=0$, which, from the expression given in (\ref{app5}), will imply its convergence to the fixed point (\ref{app3}).

Finally, from the asymptotic condition at (\ref{app2a}), valid for any solution defined in the positive octant ${\R}^3_+$, we will have that these solutions will also converge asymptotically to the fixed point (\ref{A23}). Therefore, we demonstrate that this fixed point is globally stable and attractor for all solutions of the system (\ref{eq2}) that have an initial condition in the positive octant.

\subsection{Global Stability of The Fixed Point in (\ref{new3})}

Using again the theory developed in \cite{redheffer}, we can define the following Lyapunov function for the fixed point (\ref{new3}) of the Lotka-Volterra system given in (\ref{new2}):
\begin{equation}
\label{app6}
V=X_1^{**}\left[\left(\frac{X_1}{X_1^{**}}\right)-\ln \left(\frac{X_1}{X_1^{**}}\right)-1\right]
+\frac{\alpha_2}{\beta_1}X_2^{**}\left[\left(\frac{X_2}{X_2^{**}}\right)-\ln \left(\frac{X_2}{X_2^{**}}\right)-1\right],
\end{equation}  
The time derivative of the function $V$ in (\ref{app6}) will be calculated as:
\begin{equation}
\label{app7}
\frac{dV}{dt}=-\left[\left(1-\alpha_1\right)\left(X_1-X_1^*\right)^2+\frac{\alpha_2({\bar k}+\beta_2)}{\beta_1}\left(X_2-X_2^*\right)^2\right]\leq 0,
\end{equation}
where we are considering that
\begin{equation}
\label{app8}
{\bar k}>-\beta_2.
\end{equation}

Analogously to what was discussed in the previous subsection, we can conclude that every solution of the ODE system (\ref{new2}), with an initial condition in $t_0=0$ such that $X_1(0)>0$ and $X_2 (0)>0$, will be defined for all time $t>0$ and will converge asymptotically to the fixed point (\ref{new3}) when $t\rightarrow\infty$, that is, this fixed point will be globally stable and attractor for all solutions of the system (\ref{new2}) contained in the positive orthant ${\R}^2_+$.

Obviously, the above conclusion will only be valid if the fixed point (\ref{new3}) exists in the positive orthant $\R^2_+$, that is, we have to impose $X_1^{**}>0$ and $X_2^{**}>0 $. It is quite straightforward to obtain that this condition of existence will be satisfied only if
\begin{equation}
\label{app9}
{\bar k}>-\frac{\beta_2(1-\alpha_1)+\beta_1\alpha_2}{1-\alpha_1+{\bar b}\alpha_2}.
\end{equation}
Therefore, we can conclude that the fixed point (\ref{new3}) exists and is globally stable if the conditions (\ref{app8}) and (\ref{app9}) are satisfied.

\end{document}